\PassOptionsToPackage{unicode}{hyperref}
\PassOptionsToPackage{hyphens}{url}
\PassOptionsToPackage{dvipsnames,svgnames,x11names}{xcolor}
\documentclass[11pt]{article}

\usepackage{booktabs}
\usepackage{float}
\usepackage{bbm}

\usepackage{amsmath,amssymb}
\usepackage{bm}
\usepackage{iftex}
\ifPDFTeX
 \usepackage[T1]{fontenc}
 \usepackage[utf8]{inputenc}
 \usepackage{textcomp}
\else
 \usepackage{unicode-math}
 \defaultfontfeatures{Scale=MatchLowercase}
 \defaultfontfeatures[\rmfamily]{Ligatures=TeX,Scale=1}
\fi
\ifPDFTeX\else 
\fi
\IfFileExists{upquote.sty}{\usepackage{upquote}}{}
\makeatletter
\@ifundefined{KOMAClassName}{%
 \IfFileExists{parskip.sty}{%
 \usepackage{parskip}
 }{%
 \setlength{\parindent}{0pt}
 \setlength{\parskip}{6pt plus 2pt minus 1pt}}
}{%
 \KOMAoptions{parskip=half}}
\makeatother
\usepackage{xcolor}
\makeatletter
\ifx\paragraph\undefined\else
 \let\oldparagraph\paragraph
 \renewcommand{\paragraph}{
 \@ifstar
 \xxxParagraphStar
 \xxxParagraphNoStar
 }
 \newcommand{\xxxParagraphStar}[1]{\oldparagraph*{#1}\mbox{}}
 \newcommand{\xxxParagraphNoStar}[1]{\oldparagraph{#1}\mbox{}}
\fi
\ifx\subparagraph\undefined\else
 \let\oldsubparagraph\subparagraph
 \renewcommand{\subparagraph}{
 \@ifstar
 \xxxSubParagraphStar
 \xxxSubParagraphNoStar
 }
 \newcommand{\xxxSubParagraphStar}[1]{\oldsubparagraph*{#1}\mbox{}}
 \newcommand{\xxxSubParagraphNoStar}[1]{\oldsubparagraph{#1}\mbox{}}
\fi
\makeatother

\usepackage{longtable,booktabs,array}
\usepackage{multirow}
\usepackage{calc}
\usepackage{etoolbox}
\makeatletter
\patchcmd\longtable{\par}{\if@noskipsec\mbox{}\fi\par}{}{}
\makeatother
\IfFileExists{footnotehyper.sty}{\usepackage{footnotehyper}}{\usepackage{footnote}}
\makesavenoteenv{longtable}
\usepackage{graphicx}
\makeatletter
\def\maxwidth{\ifdim\Gin@nat@width>\linewidth\linewidth\else\Gin@nat@width\fi}
\def\maxheight{\ifdim\Gin@nat@height>\textheight\textheight\else\Gin@nat@height\fi}
\makeatother
\setkeys{Gin}{width=\maxwidth,height=\maxheight,keepaspectratio}
\makeatletter
\def\fps@figure{htbp}
\makeatother

\makeatletter
\@ifpackageloaded{caption}{}{\usepackage{caption}}
\AtBeginDocument{%
\ifdefined\contentsname
 \renewcommand*\contentsname{Table of contents}
\else
 \newcommand\contentsname{Table of contents}
\fi
\ifdefined\listfigurename
 \renewcommand*\listfigurename{List of Figures}
\else
 \newcommand\listfigurename{List of Figures}
\fi
\ifdefined\listtablename
 \renewcommand*\listtablename{List of Tables}
\else
 \newcommand\listtablename{List of Tables}
\fi
\ifdefined\figurename
 \renewcommand*\figurename{Figure}
\else
 \newcommand\figurename{Figure}
\fi
\ifdefined\tablename
 \renewcommand*\tablename{Table}
\else
 \newcommand\tablename{Table}
\fi
}
\@ifpackageloaded{float}{}{\usepackage{float}}
\floatstyle{ruled}
\@ifundefined{c@chapter}{\newfloat{codelisting}{h}{lop}}{\newfloat{codelisting}{h}{lop}[chapter]}
\floatname{codelisting}{Listing}

\makeatother
\makeatletter
\@ifpackageloaded{caption}{}{\usepackage{caption}}
\@ifpackageloaded{subcaption}{}{\usepackage{subcaption}}
\makeatother
\usepackage{enumitem}

\ifLuaTeX
 \usepackage{selnolig}
\fi
\usepackage[]{natbib}
\usepackage{bookmark}

\IfFileExists{xurl.sty}{\usepackage{xurl}}{}
\hypersetup{
 pdftitle={Title},
 pdfauthor={Author 1; Author 2},
 pdfkeywords={3 to 6 keywords, that do not appear in the title},
 colorlinks=true,
 linkcolor={blue},
 filecolor={Maroon},
 citecolor={Blue},
 urlcolor={Blue},
 pdfcreator={LaTeX via pandoc}}

\newcommand{\anon}{1}
\newcommand{\btheta}{\boldsymbol{\theta}}
\newcommand{\CrI}{\mathrm{CrI}}

\begin{document}

\def\spacingset#1{\renewcommand{\baselinestretch}%
{#1}\small\normalsize} \spacingset{1}

\if1\anon
{
 \title{\bf A Practical Framework for Sensitivity Analysis in Externally 
Controlled Trials: An Illustration with a Bayesian Hybrid Evidence Synthesis 
Case Study}
 \author{
Xuemin Gu\thanks{Email: xgu@relaytx.com. The work was done while the author was an employee at the AbbVie.}\hspace{.2cm} ,
 Kitty Guo,
 and Jane Zhang
}
\date{15 June 2026}      
 \maketitle
} \fi

\if0\anon
{
 \bigskip
 \bigskip
 \bigskip
 \begin{center}
 {\LARGE\bf Title}
\end{center}
 \medskip
} \fi

\bigskip
\begin{abstract}

Externally controlled trials (ECTs), including single-arm studies augmented with historical data and hybrid randomized designs with partial external augmentation, are increasingly used when concurrent randomized controls are infeasible or unethical. Regulatory guidance from the FDA, EMA, and NMPA calls for sensitivity analysis of borrowing assumptions, yet provides no structured template for which analyses to run or how to interpret them together.

We propose a three-pillar framework organized around three questions: was the borrowing appropriate, did it contribute meaningful value, and are the conclusions robust to perturbation? The framework comprises eight modular analyses covering heterogeneity diagnostics, source influence, no-borrowing references, effective sample size, prior sensitivity, tipping points, alternative borrowing methods, and structural model sensitivity. It is method-agnostic and applies to both Bayesian and frequentist borrowing in patient-level or hybrid settings.

We illustrate the framework using simulated data that mimic a hybrid evidence synthesis from a historical approval of ethnic-bridging submission under a real-world-evidence regulatory pathway. That original analysis combined individual patient data from a global pivotal study and a regional real-world study with aggregate data from two published cohorts, fitted via a Bayesian longitudinal model with ethnic-difference parameters. The worked example provides a reproducible template for sensitivity analysis in ECT submissions.

\end{abstract}

\noindent%
\vfill

\newpage
\spacingset{1}

\section{Introduction}
\label{sec-intro}

\subsection{Externally Controlled Trials and the Credibility Challenge}
\label{sec:intro-ect}

Externally controlled trials (ECTs) are gaining popularity in drug development for rare diseases, pediatric indications, severe or life-threatening conditions, and other settings where concurrent randomized controls are infeasible or unethical. In these scenarios, sponsors often either use external data as historical controls in single-arm trials or supplement randomized control arms with data from prior trials, registries, electronic health records, or published literature. Regulatory agencies, including the FDA, EMA, and China's NMPA, are increasingly recognizing ECTs as valid under specific conditions, while emphasizing that their credibility depends on well-justified borrowing assumptions and the robustness of conclusions to reasonable alternative analyses \citep{fda_rwe, fda_bayesian, ich_e10, nmpa_rwd, ema_rwdqf}.

The primary statistical challenge in ECTs is information borrowing: determining how much external data should inform treatment effect estimates in the target population. Bayesian dynamic borrowing methods --- power priors \citep{Ibrahim2015}, commensurate priors \citep{Hobbs2011}, and robust meta-analytic-predictive priors \citep{Schmid2014} --- provide principled mechanisms for discounting external sources according to their compatibility with the current study. In contrast, frequentist and causal-inference approaches include propensity score integration \citep{liEfficiency2023}, doubly robust estimators \citep{Valancius2024}, test-then-pool procedures \citep{Viele2014, Yuan2019}, and newer selective borrowing methods such as Adaptive Lasso Selective Borrowing \citep{Gao2025} and Conformal Selective Borrowing \citep{Zhu2025}. Each family addresses the borrowing problem through different mathematical machinery, yet all rely on assumptions that cannot be fully verified from the data alone.

A practical complication relevant to all these methods is that external data rarely come in a uniform format. Some sources provide individual patient-level data, while others are limited to aggregate summaries extracted from published figures. Integrating both types of data in a single analysis, known as hybrid evidence synthesis, is especially common in global registration studies and rare-disease programs, where the only available external evidence often consists of published summaries. Hybrid synthesis introduces additional uncertainties, such as digitization error, ecological bias from aggregation, and uncontrolled heterogeneity in study design, that do not arise when all data are available at the patient level \citep{Ravva2014AggregateIPD}.

\subsection{The Sensitivity-Analysis Gap}
\label{sec:intro-gap}

Regulatory guidance requires sensitivity analyses of borrowing assumptions in ECTs but provides limited direction on which analyses to perform or how to organize and interpret them. The methods literature has focused primarily on developing individual borrowing techniques rather than on structuring a comprehensive credibility assessment. Consequently, study teams have access to sophisticated borrowing methods but limited practical guidance on assembling a robust package of sensitivity analyses. In practice, the selection of sensitivity analyses is largely left to study teams, while their weighting and interpretation are determined by regulatory reviewers. This ad-hoc approach makes cross-submission comparisons difficult. What is needed is a structured, method-agnostic framework for evaluating any chosen borrowing approach.

\subsection{Contribution and Worked Example}
\label{sec:intro-contribution}

This paper proposes a three-pillar sensitivity analysis framework organized around the questions of appropriateness, value, and robustness, decomposed into eight modular analyses that cover heterogeneity diagnostics, source influence, no-borrowing references, effective sample size, prior sensitivity, tipping points, alternative borrowing methods, and structural model sensitivity. The framework is method-agnostic: it applies regardless of whether the primary borrowing mechanism is Bayesian or frequentist, regardless of whether external data are patient-level or aggregate, and regardless of whether the design is a single-arm trial with full external control substitution or a hybrid randomized trial with partial augmentation. It is intended to complement the method literature by providing a structured way to evaluate any primary borrowing analysis method.

We illustrate the framework with a worked example that mirrors the hybrid evidence packages supporting a regulatory approval with real-world-evidence (RWE). The original hybrid evidence synthesis combined individual patient data from a global pivotal study and a regional real-world study with aggregate data from two published papers. All four sources were jointly analyzed under a Bayesian longitudinal model with explicit ethnic-difference parameters. Our worked example follows the same structure: two IPD sources, two AD sources, an Asian-versus-non-Asian comparison of a continuous longitudinal outcome, and the same modeling family. All data are simulated. Although the inferential target is a population-difference parameter rather than a treatment effect, the key challenges such as the borrowing operation, source heterogeneity, the hybrid IPD-plus-AD data structure, and the associated regulatory credibility questions are precisely those faced in ECT submissions.

\subsection{Paper Organization}
\label{sec:intro-org}

Section~\ref{sec:framework} presents the three-pillar framework in general terms applicable to externally controlled trials. Section~\ref{sec:casestudy} describes the data sources and Bayesian longitudinal model used in the worked example. Section~\ref{sec:implementation} applies the framework by conducting all eight sensitivity analyses on the worked example. Section~\ref{sec:discussion} summarizes the lessons learned, provides practical recommendations for ECT submissions, discusses the complementarity of the framework with methods offering strong internal-validity guarantees, acknowledges limitations of the framework, and offers concluding remarks. Complete mathematical specifications for all models and sensitivity analyses are provided in the Supplementary Appendix.

\section{A Three-Pillar Framework for Sensitivity Analysis in Information Borrowing}
\label{sec:framework}

Any team that borrows external information must address three fundamental questions. First, was the borrowing appropriate—are the sources sufficiently compatible and does any single source shows dominant influence? Second, what value did borrowing contribute—would the conclusions hold, or even be obtainable, without the external data? Third, are the conclusions robust—do they survive changes in prior specifications, perturbations to the external data, alternative borrowing methods, and alternative outcome models? These three questions form the organizing pillars of the sensitivity analysis framework.

\subsection{Overview}

Table~\ref{tab:framework} summarizes the framework. Each pillar is decomposed into specific analyses, labeled S1 through S8, along with the guiding question, the type of output produced, and the key analytical tools. The analyses are designed to be modular: not every study will require all eight, and practitioners should select those most relevant to their setting. However, each pillar should be represented by at least one analysis for the assessment to be considered complete.

\begin{table}[ht!]
\centering
\caption{Three-pillar sensitivity analysis framework for evaluating information borrowing.}
\label{tab:framework}
\small
\setlength{\tabcolsep}{4pt} 
\begin{tabular}{>{\raggedright\arraybackslash}p{1.0cm} 
 >{\raggedright\arraybackslash}p{3.6cm} 
 >{\raggedright\arraybackslash}p{4.0cm} 
 >{\raggedright\arraybackslash}p{3.8cm}}
\toprule
& Analysis Question & Analytical Tools & Key Output \\
\midrule
\multicolumn{4}{l}{\textbf{Pillar 1: Appropriateness of Borrowing} 
--- \textit{Was combining these sources defensible?}} \\
\addlinespace[3pt]
S1 & Are the data sources sufficiently compatible to be modeled jointly? & Predictive checks by source, forest plots, heterogeneity measures ($I^2$, $\tau^2$) & Model fit diagnostics per source \\
\addlinespace[3pt]
S2 & Does any single source dominate the conclusion? & Leave-one-source-out: refit model $K$ times, each dropping one source & Shift in primary estimate per omission \\
\midrule
\multicolumn{4}{l}{\textbf{Pillar 2: Value of Borrowing} 
--- \textit{What was gained by incorporating external data?}} \\
\addlinespace[3pt]
S3 & Would the conclusion hold without external data? & No-borrowing reference: fit model on current-study IPD only & Interval estimate widths with vs.\ without borrowing \\
\addlinespace[3pt]
S4 & How much information did each source contribute? & Effective sample size (ESS) decomposition & ESS per external source \\
\midrule
\multicolumn{4}{l}{\textbf{Pillar 3: Robustness of Conclusions} 
--- \textit{Are the conclusions stable under perturbation?}} \\
\addlinespace[3pt]
S5 & Are conclusions driven by analyst-set borrowing controls, or by the data? & Refit under alternative settings of the inputs that govern borrowing strength within the primary framework (e.g., power-prior weight $a_0$; priors on heterogeneity or comparability parameters) & Conclusion sensitivity to borrowing-control settings \\
\addlinespace[3pt]
S6 & How different would the external data need to be before the conclusion changes? & Tipping point analysis: systematically shift external data & Reversal threshold \\
\addlinespace[3pt]
S7 & Does the borrowing framework affect the conclusion? & Refit under alternative borrowing frameworks (e.g., commensurate prior, robust MAP, test-then-pool, propensity-score integration, or selective borrowing) & Comparison of primary estimates across frameworks \\
\addlinespace[3pt]
S8 & Does the outcome model affect the conclusion? & Fit alternative structural models (e.g., piecewise linear, spline) & Comparison of primary estimates across models \\
\bottomrule
\end{tabular}
\end{table}

\subsection{Pillar 1: Appropriateness of Borrowing}

The first pillar asks whether the decision to combine data sources was defensible. This is a prerequisite for the entire analysis: if the sources are fundamentally incompatible, no amount of methodological sophistication can rescue the conclusions. Appropriateness is assessed through two complementary analyses. Heterogeneity diagnostics (S1) evaluate whether the data sources are sufficiently compatible to be modeled jointly. Per-source predictive checks provide a natural tool: the fitted model generates predicted summary statistics for each source, and systematic discrepancies between predicted and observed values flag sources the model accommodates poorly. The Bayesian implementation is the posterior predictive check; in a frequentist setting, fitted-value or residual diagnostics serve the same role. Formal heterogeneity measures such as \(I^2\) (percentage of variability due to heterogeneity rather than chance) and \(\tau^2\) (between-source variance) can supplement visual diagnostics when the number of sources permits meaningful estimation. The goal is not to require homogeneity---some degree of between-source variation is expected and can be modeled explicitly---but rather to identify heterogeneity that is large enough to call the joint modeling assumption into question.

Source influence diagnostics (S2) address a related but distinct concern: even if all sources are compatible in a global sense, a single source may dominate the results to such a degree that the conclusion is effectively driven by that one dataset. A leave-one-source-out analysis---refitting the model $K$ times, each time dropping one external source---reveals which sources carry the most weight. Sources with outsized influence warrant careful scrutiny and should be the primary targets of the robustness analyses in Pillar~3.

\subsection{Pillar 2: Value of Borrowing}

The second pillar quantifies what was gained by incorporating external evidence. This is particularly important in regulatory settings, where reviewers need to understand whether the conclusions are genuinely supported by the totality of evidence or are an artifact of importing external precision into an otherwise uninformative dataset.

The no-borrowing reference (S3) provides the most direct answer: the model is fit using only the internal individual patient data (IPD), and the resulting interval estimates are compared with those from the full-borrowing analysis. If the IPD-only analysis produces intervals so wide that no meaningful conclusion can be drawn, this demonstrates that borrowing was not merely helpful but essential. Conversely, if the IPD-only analysis already supports the same conclusion, the contribution of borrowing is limited to incremental precision, and the team can be more confident that the result is not an artifact of the external data.

Effective sample size decomposition (S4) complements the no-borrowing reference by translating the precision contributed by each external source into an intuitive metric: the number of individual patients that would provide equivalent information. This is especially valuable for communicating with non-statistical audiences, including regulatory reviewers and clinical experts, who may find interval-width comparisons abstract but can readily interpret a statement such as ``the aggregate data from Publication~1 contributed the equivalent of approximately $N$ additional patients.''

\subsection{Pillar 3: Robustness of Conclusions}

The third pillar asks whether the conclusions survive perturbation. Even if borrowing was appropriate and valuable, the analysis is credible only if its conclusions do not hinge on a specific set of modeling choices or data sources. This pillar is the most analytically demanding, comprising four analyses that probe different dimensions of robustness.

Borrowing-control sensitivity (S5) tests whether conclusions are driven by analyst-chosen input paramters that control the strength of borrowing rather than by the data, holding the borrowing framework fixed. Every borrowing method has such inputs: in Bayesian dynamic borrowing these include the priors for hyperparameters, the discount factor $a_0$ in power priors, the heterogeneity scale in robust MAP priors, and priors on commensurability or heterogeneity parameters; in frequentist approaches they include equivalence margins in test-then-pool procedures, regularization parameters in selective borrowing, and trimming thresholds in propensity-score methods \citep{Viele2014, Yuan2019, Gao2025, Zhu2025, liEfficiency2023}. Refitting under plausible variation in these inputs shows whether the data alone support the conclusion. This analysis is particularly important when the chosen settings encode strong borrowing assumptions.

Tipping point analysis (S6) takes a different approach: rather than varying the model, it varies the data. The external data from the most influential source (identified in S2) are systematically shifted, and the team identifies the magnitude of perturbation required to reverse the primary conclusion. A large tipping point indicates that the conclusion is robust against potential errors in the external data, including digitization error and unmeasured heterogeneity. A small tipping point signals fragility and should be reported transparently.

Borrowing mechanism sensitivity (S7) asks whether substituting an entirely different borrowing framework changes the conclusion. Whereas S5 perturbs the controls of the primary framework, S7 substitutes a different mechanism for combining the external sources with the current-study data. The literature offers a range of alternative frameworks that include Bayesian dynamic borrowing priors \citep{Ibrahim2015, Hobbs2011, Schmid2014}, frequentist test-then-pool procedures \citep{Viele2014, Yuan2019}, propensity-score integration \citep{liEfficiency2023}, doubly robust estimators \citep{Valancius2024}, and selective borrowing approaches \citep{Gao2025, Zhu2025}. These frameworks differ in how they parameterize the degree of borrowing and how they respond to between-source discrepancy. If the primary analysis used one framework, refitting under at least one alternative provides evidence that the conclusion is not an artifact of the specific borrowing mechanism. To isolate the effect of the borrowing framework, the alternative should preserve the primary analysis's core modeling structure and differ only in the borrowing mechanism itself. In practice, this constraint narrows the realistic choice of S7 alternatives to frameworks that share the primary analysis's likelihood and outcome model: a Bayesian primary analysis is most informatively compared with alternative Bayesian frameworks, and a frequentist primary analysis with alternative frequentist or causal-inference frameworks. Crossing the Bayesian--frequentist boundary in S7 would conflate changes in the borrowing mechanism with changes in the inferential framework and the outcome model, defeating the analysis's purpose.

Finally, structural model sensitivity (S8) tests whether the parametric model for the outcome influences the conclusion. In longitudinal settings, the choice between a parametric model, a piecewise linear model, or a spline-based model may affect the estimated treatment effect or ethnic difference, particularly if the data are sparse at certain time points. Fitting at least one alternative structural model ensures that the conclusion is not an artifact of functional form assumptions.

\subsection{Interpretation and Communication}

We advocate for a three-part conclusion template in which the study team reports that borrowing was (a)~appropriate, (b)~materially valuable, and (c)~led to robust conclusions---or, where the evidence warrants, reports transparently on which of these conditions was not fully satisfied. Inconsistent results across sensitivity analyses should not be suppressed; rather, they reveal which assumptions are most critical to the conclusion and allow regulators to make informed judgments about the strength of the evidence. The goal of the framework is not to guarantee a favorable outcome but to ensure that the evaluation of information borrowing is systematic, complete, and honest.

\section{Worked Example: Hybrid Evidence Synthesis}
\label{sec:casestudy}

\subsection{Overview}
\label{sec:casestudy-overview}

We illustrate the framework on a worked example that mirrors the structure of a previous ethnic-bridging submissions under regulatory real-world evidence pathways. The setting is a chronic, progressive condition of unknown etiology, in which a continuous pharmacodynamic (PD) biomarker serves as a key manifestation of the disease. Reduction of this PD biomarker after treatment is expected to slow disease progression. The biomarker is measured longitudinally before and after a one-time intervention, and the inferential target is whether the biomarker response trajectory differs between Asian and non-Asian patients. The biomarker response is quantified and modeled as the proportion change from baseline; we refer to this measure as the clinical score, with negative values indicating reductions from baseline.

The remainder of this section describes the data sources, statistical model, and primary results of the ethnic-bridging analysis. The primary analysis combines IPD from the Global study and the RWE study with aggregate data from Publication 1 and Publication 2, and fits a Bayesian longitudinal Emax model with linear trend and explicit ethnic-difference parameters. Retrospective application of the three-pillar framework is provided in Section~\ref{sec:implementation}.

\subsection{Data Sources}
\label{sec:data}

The worked example combines four simulated data sources to represent a classic hybrid evidence synthesis setting: two sources provide individual patient data (IPD) and two provide aggregate summaries (AD). These sources are constructed to be representative of the actual submission packages in which hybrid synthesis was used. To match real-world scenarios, heterogeneity is intentionally introduced in sample sizes, baseline biomarker values, follow-up schedules, and reporting formats that the statistical model must accommodate. The summaries of biomarker values and the clinical score (proportion change from baseline) are given in Tables~\ref{tab:p13001}--\ref{tab:pub2} for each data source. For the Global study, the summary is provided for Asian and non-Asian patients separately.

\subsubsection{Global Study (IPD).}
The simulated Global pivotal study provides IPD for a total of 100 patients (90 non-Asian patients and 10 Asian patients). The mean baseline values of PD biomarker are 24.23 (non-Asian) and 20.46 (Asian). Follow-up data spans ten longitudinal visits over 336 days, with post-intervention measurements evaluated at nine specific time points: Days 1, 7, 14, 28, 84, 168, 224, 280, and 336.

\subsubsection{RWE Study (IPD).}
A simulated regional real-world evidence study provides IPD for 50 Asian patients with a study design and clinical setting highly comparable to the Global study. Individual patient data are captured across eight post-intervention visit time points over 336 days: Days 1, 7, 28, 84, 168, 224, 280, and 336. The group exhibits a mean baseline biomarker value of 24.26.

\subsubsection{Publication 1 (Aggregate Data).}
The first aggregate-data source represents a published cohort summary of Asian population, utilized here as anchoring external evidence. It provides a retrospective summary of 60 Asian patients, reporting mean and standard deviations of both the raw biomarker values and the clinical scores at five distinct post-intervention time points (Days 1, 7, 28, 84, and 168). The simulated mean baseline biomarker value is 22.0, which is notably lower than the corresponding Global study population.

\subsubsection{Publication 2 (Aggregate Data).}
The second aggregate-data source serves as another layer of anchoring external evidence for Asian population with aggregated data only. It is simulated to represent a prospective summary of 40 Asian patients, reporting aggregate clinical data across six explicit longitudinal time points: Days 1, 7, 28, 84, 168, and 336. The baseline mean biomarker value for this cohort is 26.0.

\begin{table}[ht!]
\centering
\caption{Summary of biomarker and clinical score (proportion change from baseline) in the simulated Global study.}
\label{tab:p13001}
\small
\begin{tabular}{rrrrrrrrrr}
\toprule
 & \multicolumn{2}{c}{N} & \multicolumn{2}{c}{Mean biomarker} & \multicolumn{4}{c}{Mean clinical score} \\
\cmidrule(lr){2-3} \cmidrule(lr){4-5} \cmidrule(lr){6-9}
Day & Non-Asian & Asian & Non-Asian & Asian & Mean (NA) & SE (NA) & Mean (A) & SE (A) \\
\midrule
0 & 90 & 10 & 24.23 & 20.46 & -- & -- & -- & -- \\
1 & 86 & 10 & 8.48 & 7.70 & $-$0.640 & 0.007 & $-$0.613 & 0.024 \\
7 & 89 & 8 & 12.59 & 11.69 & $-$0.475 & 0.008 & $-$0.435 & 0.019 \\
14 & 86 & 10 & 14.87 & 12.64 & $-$0.376 & 0.007 & $-$0.375 & 0.017 \\
28 & 88 & 10 & 16.12 & 14.13 & $-$0.328 & 0.006 & $-$0.300 & 0.021 \\
84 & 79 & 10 & 17.18 & 15.48 & $-$0.286 & 0.007 & $-$0.229 & 0.024 \\
168 & 79 & 10 & 17.02 & 14.60 & $-$0.296 & 0.007 & $-$0.277 & 0.020 \\
224 & 86 & 9 & 16.22 & 14.19 & $-$0.320 & 0.007 & $-$0.276 & 0.030 \\
280 & 82 & 9 & 16.26 & 14.15 & $-$0.325 & 0.007 & $-$0.323 & 0.023 \\
336 & 83 & 9 & 16.00 & 13.25 & $-$0.337 & 0.006 & $-$0.314 & 0.027 \\
\bottomrule
\end{tabular}
\end{table}

\begin{table}[ht!]
\centering
\caption{Summary of biomarker and clinical score (proportion change from baseline) in the simulated RWE study.}
\label{tab:rwe}
\small
\begin{tabular}{rrrrrrrr}
\toprule
Day & N & Mean biomarker & SE & Mean $\Delta$ & SE of $\Delta$ & Mean clinical score & SE of score \\
\midrule
0 & 50 & 24.26 & 1.29 & -- & -- & -- & -- \\
1 & 50 & 7.28 & 0.25 & $-$16.98 & 1.17 & $-$0.666 & 0.015 \\
7 & 48 & 11.48 & 0.40 & $-$12.27 & 0.97 & $-$0.479 & 0.016 \\
28 & 50 & 15.33 & 0.59 & $-$8.93 & 0.76 & $-$0.331 & 0.016 \\
84 & 43 & 16.27 & 0.70 & $-$7.81 & 0.74 & $-$0.290 & 0.016 \\
168 & 47 & 16.50 & 0.70 & $-$8.16 & 0.71 & $-$0.300 & 0.015 \\
224 & 48 & 15.63 & 0.64 & $-$8.40 & 0.80 & $-$0.311 & 0.017 \\
280 & 46 & 15.32 & 0.63 & $-$8.77 & 0.83 & $-$0.325 & 0.016 \\
336 & 43 & 15.19 & 0.65 & $-$9.38 & 0.86 & $-$0.350 & 0.016 \\
\bottomrule
\end{tabular}
\end{table}

\begin{table}[ht!]
\centering
\caption{Aggregate data from the simulated Publication~1 source.}
\label{tab:pub1}
\small
\begin{tabular}{rrrrrrrrr}
\toprule
Day & N & SD & Mean biomarker & SE & Mean $\Delta$ & SE of $\Delta$ & Mean clinical score & SE of score \\
\midrule
0 & 60 & -- & 22.00 & -- & 0.00 & -- & 0.00 & -- \\
1 & 60 & -- & 8.07 & -- & $-$13.93 & -- & $-$0.633 & -- \\
7 & 60 & -- & 11.90 & -- & $-$10.10 & -- & $-$0.459 & -- \\
28 & 60 & -- & 15.32 & -- & $-$6.68 & -- & $-$0.304 & -- \\
84 & 60 & -- & 16.39 & -- & $-$5.61 & -- & $-$0.255 & -- \\
168 & 60 & -- & 16.08 & -- & $-$5.92 & -- & $-$0.269 & -- \\
\bottomrule
\end{tabular}
\end{table}

\begin{table}[ht!]
\centering
\caption{Aggregate data from the simulated Publication~2 source.}
\label{tab:pub2}
\small
\begin{tabular}{rrrrrrrrr}
\toprule
Day & N & SD & Mean biomarker & SE & Mean $\Delta$ & SE of $\Delta$ & Mean clinical score & SE of score \\
\midrule
0 & 40 & -- & 26.00 & -- & 0.00 & -- & 0.00 & -- \\
1 & 40 & -- & 8.53 & -- & $-$17.47 & -- & $-$0.672 & -- \\
7 & 40 & -- & 12.74 & -- & $-$13.26 & -- & $-$0.510 & -- \\
28 & 40 & -- & 17.42 & -- & $-$8.58 & -- & $-$0.330 & -- \\
84 & 40 & -- & 18.03 & -- & $-$7.97 & -- & $-$0.306 & -- \\
168 & 40 & -- & 17.66 & -- & $-$8.34 & -- & $-$0.321 & -- \\
336 & 40 & -- & 16.70 & -- & $-$9.30 & -- & $-$0.358 & -- \\
\bottomrule
\end{tabular}
\end{table}

\subsection{Statistical Model}

\subsubsection{Bayesian Longitudinal Emax Model}

The primary endpoint for modeling was the clinical score, defined as the proportion change of biomarker value from baseline. By using each patient’s own baseline biomarker value as the denominator in this calculation, the clinical score inherently provides partial control for heterogeneity in baseline levels across data sources. This endpoint, together with the inclusion of patient-level baseline biomarker as a covariate in the model, was chosen to further accommodate the remaining differences in baseline values. These differences arose because studies varied in pre-baseline exposure to biomarker-lowering medicines: some patients received such treatment shortly before baseline, while others underwent a washout period. Modeling baseline biomarker as a covariate enables explicit patient-level adjustment without requiring unavailable information on prior medication use (typically absent in aggregate data). This combined approach effectively controls for the additional post-baseline variation induced by differences in pre-study treatment.

A Bayesian parametric longitudinal model with two components was used to estimate the non-monotone change in the post-baseline clinical score over time. The model has two components: a sigmoidal Emax term for the rapid early response and stabilization, and a linear long-term trend. Both ethnic groups are fitted simultaneously.

Let $y_i$ denote the clinical score for row $i$ of the combined dataset, where each row corresponds either to an individual patient data (IPD) at the $i$th visit or to a visit-level summary from an aggregate data (AD). The observation model is
\begin{equation}
 y_i \;\sim\;
 \mathcal{N}\!\left(\mu_i,\;\frac{\sigma^2}{n_i}\right),
 \label{eq:XEN-obs}
\end{equation}
where $\sigma^2$ is the residual variance and $n_i$ equals 1 for individual patient observations or the reported sample size for aggregate summaries. This variance scaling ensures that each IPD observation contributes individually to the likelihood, while each aggregate summary contributes with precision proportional to its sample size. The scaling implicitly assumes full borrowing of all external sources, equivalent to $a_0=1$ in power-prior terminology. Sensitivity to this assumption is examined under Pillar~3.

Write $d_i$ for the visit index of row $i$ and $x_{d_i}$ for the corresponding study day. The mean response $\mu_i$ is a function of study day, $x_{d_i}$, at visit $i$, and takes the form
\begin{equation}
 \mu_i(x_{d_i})
 = \underbrace{E_0 + I_i\,\Delta E_0}_{\text{initial biomarker drop}}
 + \underbrace{
 \frac{
 \left(E_{\max} + I_i\,\Delta E_{\max}\right)
 x_{d_i}^{\,r + I_i\,\Delta r}
 }{
 \left(ED_{50} + I_i\,\Delta ED_{50}\right)^{r + I_i\,\Delta r}
 + x_{d_i}^{\,r + I_i\,\Delta r}
 }
 }_{\text{Emax component}}
 + \; a\,z_i
 + \underbrace{\left(b + I_i\,\Delta b\right)x_{d_i}}_%
 {\text{linear trend}},
 \label{eq:XEN-mean}
\end{equation}
where $z_i$ is the baseline biomarker value for row $i$ (the patient's baseline value for IPD, the reported source summary for AD), and $I_i \in \{0,1\}$ is an indicator equal to 1 for Asian patients and 0 for non-Asian patients. The additive delta parameters ($\Delta E_0$, $\Delta E_{\max}$, $\Delta ED_{50}$, $\Delta r$, $\Delta b$) capture potential ethnic differences in each parameter. For Asian patients ($I_i=1$), the effective parameter is, e.g., $E_0^{\text{Asian}} = E_0 + \Delta E_0$. The primary hypothesis of ethnic comparability is that all $\Delta$ parameters are close to zero. Table~\ref{tab:XEN-params} summarizes the interpretation of each parameter.

\begin{table}[ht]
\centering
\caption{Parameter definitions for the Emax longitudinal model \eqref{eq:XEN-obs}--\eqref{eq:XEN-mean}.}
\label{tab:XEN-params}
\begin{tabular}{lll}
\toprule
\textbf{Parameter} & \textbf{Interpretation} & \textbf{Applies to} \\
\midrule
$E_0$ & Immediate post-intervention reduction in the clinical score & Non-Asian \\
$E_{\max}$ & Maximum saturation value of clinical score over time & Non-Asian \\
$ED_{50}$ & Days to reach half of $E_{\max}$ from $E_0$ & Non-Asian \\
$r$ & Hill coefficient (slope at $ED_{50}$) & Non-Asian \\
$b$ & Long-term linear drift rate (per day) & Non-Asian \\
$a$ & Baseline biomarker covariate effect on $E_0$ & Both \\
\addlinespace
$\Delta(\cdot)$ & Ethnic offsets (Asian $-$ Non-Asian) for each parameter & Asian \\
\bottomrule
\end{tabular}
\end{table}

\subsubsection{Prior distributions.}
Relatively non-informative priors are specified for the main parameters, with tighter priors on the ethnic offsets to improve numerical stability:
\begin{align}
 E_0 &\sim \mathcal{N}(0, 10), &
 E_{\max} &\sim \mathcal{N}(0, 1), \nonumber \\
 ED_{50} &\sim \mathcal{N}(10, 10)\cdot\mathbf{1}(ED_{50}>0), &
 r &\sim \mathcal{N}(0, 100)\cdot\mathbf{1}(r>0), \nonumber \\
 b &\sim \mathcal{N}(0, 100), &
 a &\sim \mathcal{N}(0, 100), \nonumber \\[0.5em]
 \Delta E_0 &\sim \mathcal{N}(0, 1), &
 \Delta E_{\max} &\sim \mathcal{N}(0, 1), \nonumber \\
 \Delta ED_{50} &\sim \mathcal{N}(0, 10)\cdot\mathbf{1}(\Delta ED_{50}>-ED_{50}), &
 \Delta r &\sim \mathcal{N}(0, 100)\cdot\mathbf{1}(\Delta r>-r), \nonumber \\
 \Delta b &\sim \mathcal{N}(0, 10), &
 \sigma^2 &\sim \mathrm{InvGamma}(0.01, 0.01).
 \label{eq:prior}
\end{align}
Throughout Eq.~\eqref{eq:prior}, $\mathcal{N}(\mu, v)$ denotes the normal distribution with mean $\mu$ and variance $v$ (not standard deviation). Truncated normals are written as $\mathcal{N}(\mu, v)\cdot\mathbf{1}(\cdot)$. The $\mathrm{InvGamma}(0.01, 0.01)$ prior on the residual variance $\sigma^2$ is essentially non-informative, placing minimal constraint on the scale of unexplained variability. Prior selection followed a systematic comparison of five candidates spanning a wide range of informativeness: $\mathrm{IG}(100,1)$, $\mathrm{IG}(3, 0.20)$, $\mathrm{IG}(2, 0.15)$, $\mathrm{IG}(1,1)$, and $\mathrm{IG}(0.01, 0.01)$. All five candidates achieved satisfactory convergence in the point estimates of the Gelman--Rubin statistic ($\hat{R} < 1.05$ for all parameters). The $\mathrm{IG}(0.01, 0.01)$ prior is selected as the primary specification because it is the most diffuse candidate and imposes the least informative constraint on $\sigma^2$; its full per-parameter convergence diagnostics are reported in Table~\ref{tab:convergence}.

\subsection{Primary Results}
\label{sec:primary-results}

The Bayesian longitudinal model in equations~\eqref{eq:XEN-obs}--\eqref{eq:XEN-mean} is fitted to the combined data using all four sources at full weight. Posterior predictive curves for each ethnic group are obtained by evaluating the mean function~\eqref{eq:XEN-mean} at each study-visit day for every MCMC draw $\btheta^{(m)}$, $m=1,\dots,M$, using the group-specific mean baseline biomarker value. The posterior mean curve and 95\% credible band at each day are computed by averaging the draws and taking the 2.5\% and 97.5\% quantiles (details in Appendix~\ref{app:fitted-curves}). Overlapping credible bands between the Asian and non-Asian groups indicate comparability of the response trajectories.

\begin{figure}[ht!]
 \centering
 \includegraphics[width=0.7\textwidth]{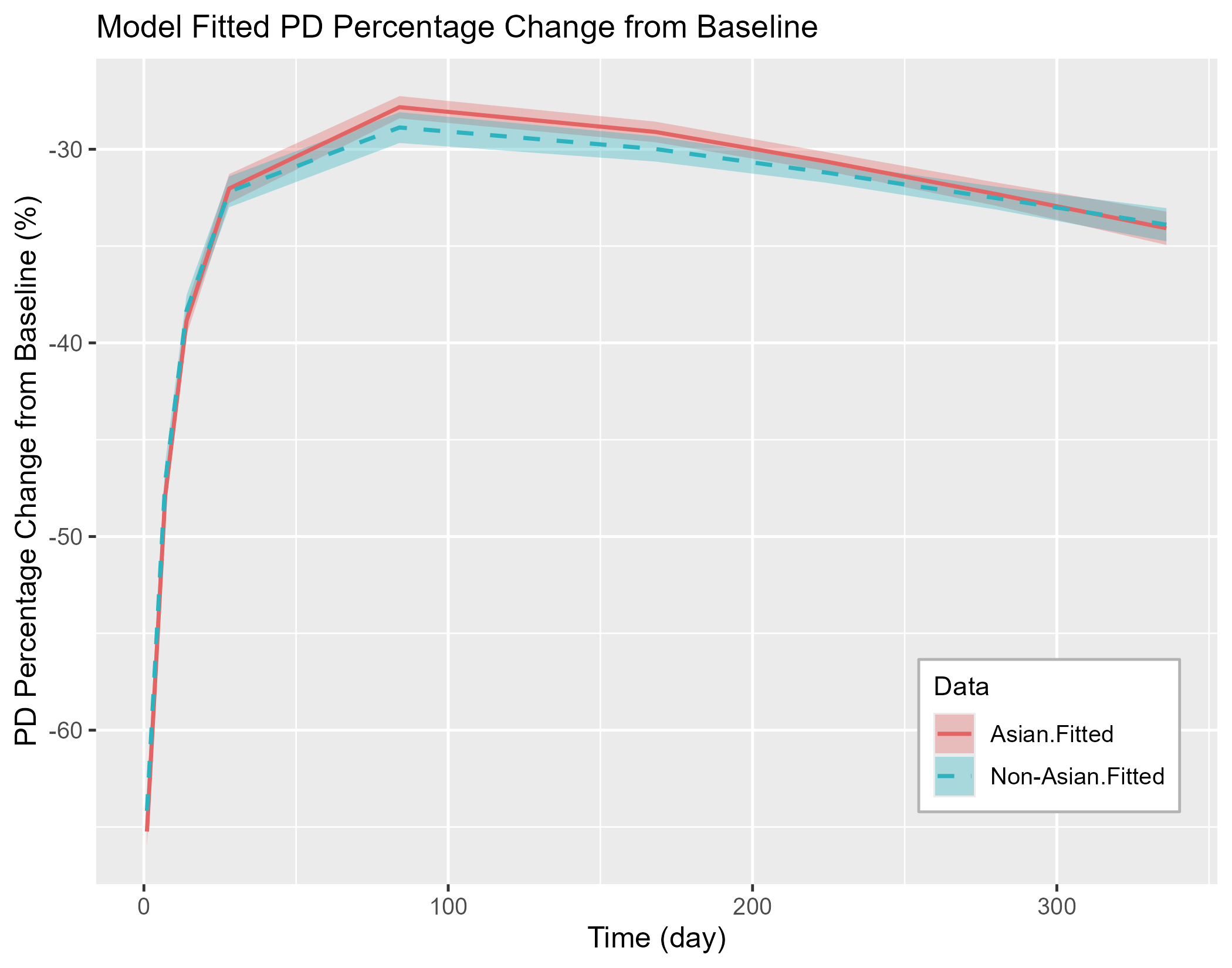}
 \caption{Comparison of model-fitted mean curves for Asian and non-Asian patients under the worked example. The two groups show a nearly identical immediate response and converge toward a similar plateau after the Emax phase, with credible bands overlapping throughout, consistent with the Asian-versus-non-Asian comparability hypothesis across all difference parameters. Fitted curves are plotted in percentage change from baseline.}
 \label{fig:comparison}
\end{figure}

Bayesian posterior summaries of all model parameters are presented in Table~\ref{tab:params}. The posterior mean of $E_0$ is $-0.668$, indicating roughly a 67\% reduction in the clinical score immediately post-intervention for non-Asian patients. The score then rebounds gradually with a maximum partial rebound of approximately $E_{\max} \approx 0.425$, half of which is achieved within approximately one week ($ED_{50} \approx 7.5$ days). The estimated long-term linear-drift coefficient $b$ is essentially zero, with the post-rebound plateau remaining stable through the end of follow-up. The posterior mean of $\Delta E_0$ is $-0.024$ with a 95\% credible interval (CrI) of $[-0.059, 0.013]$ that covers zero, providing no evidence of an ethnic difference in the immediate response. The remaining ethnic-difference parameters ($\Delta E_{\max}$, $\Delta ED_{50}$, $\Delta r$, $\Delta b$) likewise have 95\% CrIs that cover zero, providing no evidence of meaningful ethnic differences in the shape or long-term trajectory of the rebound. The primary analysis is therefore consistent with ethnic comparability across all five difference parameters, and the fitted Asian and non-Asian curves track closely throughout follow-up (Figure~\ref{fig:comparison}).

\begin{table}[ht!]
\centering
\caption{Bayesian posterior summary of model parameters for the worked example.}
\label{tab:params}
\small
\begin{tabular}{lrrrrr}
\toprule
Parameter & Mean & Median & SD & 2.5\% & 97.5\% \\
\midrule
$E_0$            & $-$0.6680 & $-$0.6665 & 0.0136 & $-$0.6995 & $-$0.6458 \\
$\Delta E_0$     & $-$0.0238 & $-$0.0241 & 0.0179 & $-$0.0586 & $\ \ $0.0130 \\
$E_{\max}$       & $\ \ $0.4246 & $\ \ $0.4222 & 0.0247 & $\ \ $0.3836 & $\ \ $0.4815 \\
$\Delta E_{\max}$& $\ \ $0.0458 & $\ \ $0.0460 & 0.0341 & $-$0.0223 & $\ \ $0.1124 \\
$ED_{50}$        & $\ \ $7.545 & $\ \ $7.557 & 0.4608 & $\ \ $6.605 & $\ \ $8.426 \\
$\Delta ED_{50}$ & $\ \ $0.667 & $\ \ $0.658 & 0.6189 & $-$0.521 & $\ \ $1.915 \\
$r$              & $\ \ $1.293 & $\ \ $1.289 & 0.1476 & $\ \ $1.009 & $\ \ $1.590 \\
$\Delta r$       & $-$0.122  & $-$0.120  & 0.1825 & $-$0.487 & $\ \ $0.229 \\
$b$              & $-$3$\times 10^{-4}$ & $-$3$\times 10^{-4}$ & $<$1$\times 10^{-4}$ & $-$3$\times 10^{-4}$ & $-$2$\times 10^{-4}$ \\
$\Delta b$       & $-$1$\times 10^{-4}$ & $-$1$\times 10^{-4}$ & 1$\times 10^{-4}$ & $-$2$\times 10^{-4}$ & $\ \ $0 \\
$a$              & $-$0.0101 & $-$0.0101 & 0.0002 & $-$0.0105 & $-$0.0096 \\
$\sigma^2$       & $\ \ $0.0025 & $\ \ $0.0025 & 0.0001 & $\ \ $0.0023 & $\ \ $0.0027 \\
\bottomrule
\end{tabular}
\end{table}

The primary analysis alone leaves important questions unanswered. Was the decision to combine these heterogeneous sources defensible? How much of the precision in the ethnic comparison was actually contributed by the external aggregate data? And would the conclusion of ethnic comparability survive under alternative modeling assumptions, different priors, or perturbations to the external data? These are precisely the questions addressed by the three-pillar sensitivity analysis framework. In the next section, we apply the framework to the worked example, implementing the complete set of sensitivity analyses.

\section{Implementing the Three-Pillar Sensitivity Analysis}
\label{sec:implementation}

This section applies the three-pillar framework to evaluate the primary analysis systematically. We proceed through each pillar in turn, presenting the approach and results of each analysis on the worked example. All sensitivity analyses of the data were conducted using \texttt{nimble} in \texttt{R} \citep{deValpine2017}.

\subsection{Pillar 1: Appropriateness of Borrowing}

\subsubsection{S1 --- Heterogeneity Diagnostics}

The four data sources differ in baseline biomarker levels (Publication 2 = 26.00 versus Global Asian = 20.46), follow-up schedules, and data format. To evaluate whether these differences are compatible with joint modeling, we conducted posterior predictive checks (PPCs). For each source and visit, posterior draws of the fitted model were compared against the observed data. For the IPD sources, summaries were aggregated at the visit level because per-patient posterior predictive $p$-values primarily reflect individual-level variability rather than overall model fit. 

Table~\ref{tab:ppc} presents, for each source-by-visit combination, the number of observations, the observed mean (or digitized value for aggregate data), the posterior predictive mean, the residual, the standard error of the source-level mean, the standardized residual $z =\mathrm{residual}/\mathrm{SE}_{\mathrm{mean}}$, and the posterior predictive $p$-value (PPP). The PPP represents the posterior probability that a model-replicated source-level mean does not exceed the observed value. Values close to 0 or 1 indicate systematic over- or under-prediction, respectively, while values near 0.5 suggest adequate model fit.

\begin{table}[ht!]
\centering
\caption{S1 --- Posterior predictive check summary by source and visit. For IPD sources (Global non-Asian, Global Asian, RWE), observations are aggregated across patients within each visit. For aggregate sources (Publication~1, Publication~2), each row corresponds to the published summary at that visit ($\mathrm{SE}_{\mathrm{mean}}$ and $z$ are not defined for aggregate observation). The standardized residual $z$ has approximate magnitude $> 2$ when the source-level mean is incompatible with the model-implied mean. The posterior predictive $p$-value (PPP) is $P(\tilde{\mu}_{sd} \le y_{sd}^{\mathrm{obs}} \mid y)$.}
\label{tab:ppc}
\small
\begin{tabular}{lrrrrrrr}
\toprule
Source & Day & $N$ & Obs.\ mean & Predicted & Residual & $z$ & $p$ \\
\midrule
Global Non-Asian & 1 & 86 & $-$0.6404 & $-$0.6417 & $\ \ $0.0013 & $\ \ $0.18 & 0.591 \\
 & 7 & 89 & $-$0.4745 & $-$0.4718 & $-$0.0027 & $-$0.35 & 0.295 \\
 & 14 & 86 & $-$0.3756 & $-$0.3835 & $\ \ $0.0078 & $\ \ $1.14 & 0.971 \\
 & 28 & 88 & $-$0.3276 & $-$0.3219 & $-$0.0057 & $-$0.88 & 0.083 \\
 & 84 & 79 & $-$0.2864 & $-$0.2888 & $\ \ $0.0024 & $\ \ $0.36 & 0.721 \\
 & 168 & 79 & $-$0.2957 & $-$0.2999 & $\ \ $0.0041 & $\ \ $0.63 & 0.889 \\
 & 224 & 86 & $-$0.3202 & $-$0.3120 & $-$0.0082 & $-$1.11 & 0.001 \\
 & 280 & 82 & $-$0.3251 & $-$0.3252 & $\ \ $0.0001 & $\ \ $0.01 & 0.513 \\
 & 336 & 83 & $-$0.3374 & $-$0.3389 & $\ \ $0.0016 & $\ \ $0.25 & 0.638 \\
\addlinespace[2pt]
Global Asian & 1 & 10 & $-$0.6128 & $-$0.6524 & $\ \ $0.0395 & $\ \ $1.65 & 1.000 \\
 & 7 & 8 & $-$0.4352 & $-$0.4789 & $\ \ $0.0436 & $\ \ $2.32 & 1.000 \\
 & 14 & 10 & $-$0.3753 & $-$0.3887 & $\ \ $0.0134 & $\ \ $0.77 & 0.999 \\
 & 28 & 10 & $-$0.3004 & $-$0.3203 & $\ \ $0.0199 & $\ \ $0.95 & 1.000 \\
 & 84 & 10 & $-$0.2289 & $-$0.2783 & $\ \ $0.0494 & $\ \ $2.04 & 1.000 \\
 & 168 & 10 & $-$0.2769 & $-$0.2910 & $\ \ $0.0142 & $\ \ $0.70 & 1.000 \\
 & 224 & 9 & $-$0.2757 & $-$0.3063 & $\ \ $0.0306 & $\ \ $1.03 & 1.000 \\
 & 280 & 9 & $-$0.3231 & $-$0.3232 & $\ \ $0.0001 & $\ \ $0.00 & 0.507 \\
 & 336 & 9 & $-$0.3141 & $-$0.3408 & $\ \ $0.0267 & $\ \ $0.99 & 1.000 \\
\addlinespace[2pt]
RWE (regional) & 1 & 50 & $-$0.6660 & $-$0.6524 & $-$0.0137 & $-$0.92 & 0.000 \\
 & 7 & 48 & $-$0.4793 & $-$0.4789 & $-$0.0005 & $-$0.03 & 0.453 \\
 & 28 & 50 & $-$0.3308 & $-$0.3203 & $-$0.0105 & $-$0.66 & 0.003 \\
 & 84 & 43 & $-$0.2902 & $-$0.2783 & $-$0.0119 & $-$0.75 & 0.000 \\
 & 168 & 47 & $-$0.2999 & $-$0.2910 & $-$0.0088 & $-$0.60 & 0.001 \\
 & 224 & 48 & $-$0.3110 & $-$0.3063 & $-$0.0047 & $-$0.28 & 0.029 \\
 & 280 & 46 & $-$0.3250 & $-$0.3232 & $-$0.0018 & $-$0.11 & 0.276 \\
 & 336 & 43 & $-$0.3499 & $-$0.3408 & $-$0.0091 & $-$0.56 & 0.019 \\
\addlinespace[2pt]
Publication~1 & 1 & 60 & $-$0.6333 & $-$0.6524 & $\ \ $0.0191 & --- & 1.000 \\
 & 7 & 60 & $-$0.4590 & $-$0.4789 & $\ \ $0.0199 & --- & 1.000 \\
 & 28 & 60 & $-$0.3038 & $-$0.3203 & $\ \ $0.0165 & --- & 1.000 \\
 & 84 & 60 & $-$0.2550 & $-$0.2783 & $\ \ $0.0233 & --- & 1.000 \\
 & 168 & 60 & $-$0.2690 & $-$0.2910 & $\ \ $0.0220 & --- & 1.000 \\
\addlinespace[2pt]
Publication~2 & 1 & 40 & $-$0.6721 & $-$0.6524 & $-$0.0197 & --- & 0.000 \\
 & 7 & 40 & $-$0.5100 & $-$0.4789 & $-$0.0311 & --- & 0.000 \\
 & 28 & 40 & $-$0.3301 & $-$0.3203 & $-$0.0098 & --- & 0.005 \\
 & 84 & 40 & $-$0.3064 & $-$0.2783 & $-$0.0282 & --- & 0.000 \\
 & 168 & 40 & $-$0.3207 & $-$0.2910 & $-$0.0296 & --- & 0.000 \\
 & 336 & 40 & $-$0.3577 & $-$0.3408 & $-$0.0169 & --- & 0.000 \\
\bottomrule
\end{tabular}
\end{table}

As show in Table~\ref{tab:ppc}, the model fits the dominant Global non-Asian source closely: every standardized residual is below 1.2 in absolute value and every mean residual is below 0.01. The single extreme posterior-predictive $p$-value, at Day~224 ($p = 0.001$), arises from an unusually small source-level standard error rather than a large residual: the residual there is only $-0.008$. The close fit to this source is expected, since the non-Asian patients in the Global study contribute the majority of the observations driving parameter estimation. Second, the Global Asian source ($n \le 10$ per visit) shows uniformly positive residuals (up to 0.05) with posterior-predictive $p$-values at or near 1 and standardized residuals as large as 2.32. This pattern reflects the small per-visit sample sizes and tight within-visit clustering, which produce source-level standard errors too small to serve as reliable denominators, rather than a structural failure of the model; the residuals themselves remain modest relative to the response level.
 
Additionally, the two aggregate sources exhibit opposite systematic biases relative to the joint model in Table~\ref{tab:ppc}. Publication~1 has uniformly positive residuals (0.016 to 0.023) at every visit, with posterior-predictive $p$-values near 1, indicating that the joint model predicts a larger biomarker reduction than Publication~1 observed; Publication~1 is systematically shallower than the joint fit. Publication~2 has uniformly negative residuals ($-$0.010 to $-$0.031) at every visit, with $p$-values near 0, indicating the opposite: the joint model predicts a smaller reduction than Publication~2 observed, so Publication~2 is systematically deeper than the joint fit. The two aggregate sources pull the posterior in opposite directions and the fitted curve splits the difference.
 
The RWE source in Table~\ref{tab:ppc} shows uniformly negative residuals ($-$0.0005 to $-$0.014) across all visits, with extreme posterior-predictive $p$-values at several of the early-to-mid visits (Days~1, 28, 84, and 168). The RWE trajectory is therefore mildly deeper than the joint model predicts. Although the magnitude of the misfit is modest, the consistent direction across visits represents precisely the type of systematic signal that posterior predictive checks are designed to detect.
 
The S1 diagnostics are consistent with joint modeling: the dominant non-Asian source is closely reproduced, the Global Asian residuals reflect limited per-visit sample sizes rather than model misspecification, and the opposing biases of Publication~1 and Publication~2 partially offset one another in the joint posterior. The diagnostics also reveal informative between-source structure that motivates the remaining analyses: the systematic opposing biases of the two aggregate sources provide natural targets for source-influence (S2), tipping-point (S6), and alternative-framework (S7) analyses, while the pattern and magnitude of source-level residuals motivate the precision decomposition in S4.

\subsubsection{S2 --- Source Influence Diagnostics}

The model was refit four times, each time dropping one Asian data source (Global Asian subgroup, RWE, Publication~1, Publication~2). The posterior distributions for the ethnic-difference parameters were compared across all five fits. Baseline biomarker centering values were recomputed for each reduced dataset to avoid introducing artificial shifts.

Table~\ref{tab:loso} summarizes the posterior means and 95\% CrIs for the primary ethnic-difference parameters across all five leave-one-source-out fits. Publication~2 is the most influential source: dropping it shifts the posterior mean of $\Delta E_0$ from $-0.024$ to $-0.043$ and moves the 95\% CrI to $[-0.081, -0.008]$, which now excludes zero, while the CrI for $\Delta E_{\max}$ shifts to $[0.017, 0.168]$ and also excludes zero. In other words, removing Publication~2 reverses the comparability conclusion for both parameters. In contrast, dropping the Global Asian subgroup, the RWE source, or Publication~1 produces only minor changes, leaving every $\Delta$ parameter CrI covering zero.

Across the four fits that retain Publication~2, the 95\% CrI for $\Delta E_0$ continues to cover zero, indicating that the comparability conclusion is stable to the omission of any single source other than Publication~2. The result is therefore contingent on the longest-follow-up aggregate source rather than robust to dropping every source. Because Publication~2 provides the only aggregate data extending through Day~336 and exerts this disproportionate influence, it is the natural target for the directed perturbation in the tipping-point probe of S6.

\begin{table}[ht!]
\centering
\caption{S2 --- Leave-one-source-out analysis: posterior means and 95\% credible intervals for the key ethnic-difference parameters. The first row (``None dropped'') reproduces the primary analysis. Baseline biomarker centering is recomputed for each reduced dataset.}
\label{tab:loso}
\small
\begin{tabular}{lrlrlrl}
\toprule
Dropped Source & $\Delta E_0$ & 95\% CrI & $\Delta E_{\max}$ & 95\% CrI & $\Delta ED_{50}$ & 95\% CrI \\
\midrule
None (primary) & $-$0.0238 & [$-$0.0586,\ 0.0130] & 0.0458 & [$-$0.0223,\ 0.1124] & 0.667 & [$-$0.521,\ 1.915] \\
Global Asian   & $-$0.0187 & [$-$0.0510,\ 0.0140] & 0.0328 & [$-$0.0300,\ 0.1010] & 0.595 & [$-$0.551,\ 1.882] \\
RWE (regional) & $-$0.0161 & [$-$0.0560,\ 0.0220] & 0.0415 & [$-$0.0290,\ 0.1190] & 0.940 & [$-$0.368,\ 2.180] \\
Publication~1  & $-$0.0193 & [$-$0.0450,\ 0.0070] & 0.0295 & [$-$0.0210,\ 0.0760] & 0.557 & [$-$0.736,\ 1.808] \\
Publication~2  & $-$0.0430 & [$-$0.0810, $-$0.0080] & 0.0874 & [\ \ 0.0170,\ 0.1680] & 0.680 & [$-$0.672,\ 2.071] \\
\bottomrule
\end{tabular}
\end{table}

\subsection{Pillar 2: Value of Borrowing}

\subsubsection{S3 --- No-Borrowing Reference}

The model was fit using only the IPD: the Global study patients (non-Asian and Asian subgroups) and the RWE patients. Without the aggregate data from Publication~1 and Publication~2, the 95\% credible bands for the Asian patient curve were compared to those from the full-borrowing analysis. Table~\ref{tab:no-borrow-params} reports the posterior summaries for the no-borrowing model; Table~\ref{tab:no-borrow-ci} compares the 95\% CrI widths for the fitted Asian biomarker curve at Month~6 and Month~12.

\begin{table}[ht!]
\centering
\caption{S3 --- Posterior summaries from the no-borrowing model (IPD only: Global study and RWE). Parameter names follow the notation in Eq.~\eqref{eq:XEN-mean}.}
\label{tab:no-borrow-params}
\small
\begin{tabular}{lrrrrr}
\toprule
Parameter & Mean & Median & SD & 2.5\% & 97.5\% \\
\midrule
$E_0$            & $-$0.6688 & $-$0.6670 & 0.0140 & $-$0.7010 & $-$0.6463 \\
$\Delta E_0$     & $-$0.0635 & $-$0.0555 & 0.0455 & $-$0.1733 & $-$0.0006 \\
$E_{\max}$       & $\ \ $0.4256 & $\ \ $0.4225 & 0.0252 & $\ \ $0.3850 & $\ \ $0.4835 \\
$\Delta E_{\max}$& $\ \ $0.1120 & $\ \ $0.0959 & 0.0880 & $-$0.0062 & $\ \ $0.3204 \\
$ED_{50}$        & $\ \ $7.520 & $\ \ $7.529 & 0.4663 & $\ \ $6.565 & $\ \ $8.416 \\
$\Delta ED_{50}$ & $\ \ $0.204 & $\ \ $0.204 & 0.9567 & $-$1.690 & $\ \ $2.090 \\
$r$              & $\ \ $1.287 & $\ \ $1.287 & 0.1483 & $\ \ $1.001 & $\ \ $1.581 \\
$\Delta r$       & $-$0.3266 & $-$0.3242 & 0.2384 & $-$0.7931 & $\ \ $0.1323 \\
$b$              & $-$3$\times 10^{-4}$ & $-$3$\times 10^{-4}$ & $<$1$\times 10^{-4}$ & $-$4$\times 10^{-4}$ & $-$2$\times 10^{-4}$ \\
$\Delta b$       & $-$1$\times 10^{-4}$ & $-$1$\times 10^{-4}$ & 1$\times 10^{-4}$ & $-$3$\times 10^{-4}$ & $\ \ $0 \\
$a$              & $-$0.0100 & $-$0.0100 & 2$\times 10^{-4}$ & $-$0.0105 & $-$0.0096 \\
$\sigma^2$       & $\ \ $0.0025 & $\ \ $0.0025 & 1$\times 10^{-4}$ & $\ \ $0.0023 & $\ \ $0.0027 \\
\bottomrule
\end{tabular}
\end{table}

\begin{table}[ht!]
\centering
\caption{S3 --- Comparison of 95\% CrI widths for the fitted Asian biomarker change from baseline curve under full borrowing (primary model) and no borrowing (IPD only). Widths are computed at Month~6 (Day~168) and Month~12 (Day~336). A width ratio $>1$ indicates that the no-borrowing analysis is less precise at that time point.}
\label{tab:no-borrow-ci}
\small
\begin{tabular}{lrrrr}
\toprule
Analysis & M6 CrI width & M12 CrI width & Ratio at M6 & Ratio at M12 \\
\midrule
Full Borrowing & 0.0106 & 0.0172 & \multirow{2}{*}{1.61$\times$} & \multirow{2}{*}{1.30$\times$} \\
No Borrowing  & 0.0171 & 0.0224 & & \\
\bottomrule
\end{tabular}
\end{table}

The no-borrowing analysis produces a posterior mean for $\Delta E_0$ of $-0.064$ with a 95\% CrI of $[-0.173, -0.001]$ that excludes zero. The comparability conclusion for $\Delta E_0$ from the primary analysis therefore depends on the inclusion of the aggregate sources: with only the IPD, the data point to a larger immediate response in Asian patients, and it is the aggregate sources that pull the posterior back across zero into comparability. The 95\% CrI for the fitted Asian curve is also 1.61 times wider at Month~6 and 1.30 times wider at Month~12 without borrowing, indicating that the aggregate sources contribute substantial precision in addition to shifting the point estimate. Borrowing in this setting therefore changes the inferential outcome rather than merely sharpening it.

\subsubsection{S4 --- Effective Sample Size Decomposition}

The precision contributed by each external source was translated into an effective sample size (ESS)---the number of individual Asian patients that would provide equivalent information. The ESS is computed by comparing the posterior precision of the fitted Asian curve with and without each source, calibrated against the per-patient precision from the IPD-only analysis (see Appendix~\ref{app:S4} for the formula).

Table~\ref{tab:ess} presents the ESS decomposition. Publication~1 contributes a substantial ESS at Month~6 (45.6 patient-equivalents) but its contribution collapses to 1.1 at Month~12, consistent with its data ending at Day~168. Publication~2 contributes 25.7 patient-equivalents at Month~6 and a sustained 27.5 at Month~12, reflecting its 336-day follow-up. The RWE source contributes 44.9 at Month~6 and 51.2 at Month~12, the largest sustained contribution across the follow-up window. The variance ratios all exceed 1, confirming that each source provides a measurable positive precision contribution.

\begin{table}[ht!]
\centering
\caption{S4 --- Effective sample size (ESS) decomposition. ESS is expressed as the number of individual Asian patients providing equivalent information. VarRatio $= V^{(-s)} / V_{\text{full}}$; values $>1$ confirm a positive precision contribution from source $s$.}
\label{tab:ess}
\resizebox{\textwidth}{!}{
\begin{tabular}{lcccccc}
\toprule
Source & $N_{\text{rows}}$ & $N_{\text{patients}}$ & ESS (M6) & ESS (M12) & VarRatio (M6) & VarRatio (M12) \\
\midrule
Publication~1  & 5   & 60 & 45.6 & $\ \ $1.1 & 1.41 & 1.01 \\
Publication~2  & 6   & 40 & 25.7 & $\ $27.5 & 1.20 & 1.37 \\
RWE (regional) & 375 & 50 & 44.9 & $\ $51.2 & 1.40 & 2.01 \\
\bottomrule
\end{tabular}
}
\end{table}

The S2 and S4 findings together clarify how each source enters the borrowing operation. S2 identified Publication~2 as the most influential single source for the location of $\Delta E_0$: dropping it moves the posterior mean from $-0.024$ to $-0.043$ and pushes the 95\% CrI across zero, reversing the comparability conclusion. S4 attributes the largest sustained precision contribution at Month~12 to RWE (ESS = 51.2) and, to a lesser extent, Publication~2 (ESS = 27.5); Publication~1's precision contribution falls sharply beyond Day~168 (ESS = 1.1 at Month~12). The two findings answer complementary questions. S2 asks where the posterior location comes from---which source holds $\Delta E_0$ within the comparability region---and Publication~2 dominates that diagnostic because its longer follow-up and deeper observed response are most strongly weighted in the joint fit. S4 asks where the posterior precision comes from, and the sources with the longest active follow-up window (RWE and Publication~2) dominate that diagnostic. Reporting both influences is necessary: Publication~2 is the location anchor that motivates the S6 tipping-point probe, while RWE and Publication~2 together are responsible for the precision gain over the no-borrowing reference observed in S3.

\subsection{Pillar 3: Robustness of Conclusions}

\subsubsection{S5 --- Borrowing-Control Sensitivity}

S5 tests whether the conclusion is driven by analyst-set inputs that control the strength of borrowing rather than by the data. The primary model implements borrowing through the variance scaling $\sigma^2/n_i$ on the aggregate sources, which is equivalent to setting the power-prior weight $a_0 = 1$ (i.e., full incorporation) for Publication~1 and Publication~2 (Equation~\eqref{eq:XEN-obs}). S5 perturbs that choice within the same power-prior family by refitting the primary model at $a_0 \in \{0.25, 0.50, 0.75, 1.00\}$. As an auxiliary check, we additionally probe the priors on the ethnic-difference parameters $\Delta$, which control how strongly the model presumes ethnic comparability across the Asian and non-Asian groups within the inference model.

\paragraph*{Power-prior weight grid.}
The aggregate-data likelihood for Publication~1 and Publication~2 is raised to the power $a_0 \in [0,1]$ while the RWE study retains full weight; the formal specification is in Appendix~\ref{app:power-prior}. Table~\ref{tab:s5-a0} reports the posterior estimates for the two primary ethnic-difference parameters across the four weights. The posterior mean of $\Delta E_0$ stays near $-0.03$ across the grid, ranging from $-0.024$ at $a_0 = 1$ to $-0.036$ at $a_0 = 0.25$, and the 95\% CrI covers zero at every weight, including the heaviest discount $a_0 = 0.25$ where the CrI is $[-0.085, 0.007]$. The CrI for $\Delta E_{\max}$ covers zero throughout. The comparability conclusion is therefore robust to the analyst's choice of borrowing strength within the primary borrowing family.

\begin{table}[ht!]
\centering
\caption{S5 --- Power-prior weight grid: posterior means and 95\% CrIs for $\Delta E_0$ and $\Delta E_{\max}$ across four discount weights $a_0$ applied to the Publication~1 and Publication~2 likelihoods. The RWE study retains full weight throughout. The row $a_0 = 1.00$ corresponds to the primary analysis (full incorporation).}
\label{tab:s5-a0}
\small
\begin{tabular}{lrlrl}
\toprule
Configuration & $\Delta E_0$ & 95\% CrI & $\Delta E_{\max}$ & 95\% CrI \\
\midrule
$a_0 = 1.00$ (primary) & $-$0.0238 & [$-$0.0586,\ 0.0130] & 0.0458 & [$-$0.0223,\ 0.1124] \\
$a_0 = 0.75$           & $-$0.0258 & [$-$0.0566,\ 0.0099] & 0.0501 & [$-$0.0037,\ 0.1087] \\
$a_0 = 0.50$           & $-$0.0245 & [$-$0.0653,\ 0.0125] & 0.0457 & [$-$0.0215,\ 0.1166] \\
$a_0 = 0.25$           & $-$0.0355 & [$-$0.0845,\ 0.0073] & 0.0653 & [$-$0.0137,\ 0.1639] \\
\bottomrule
\end{tabular}
\end{table}

\paragraph*{Ethnic-difference prior sensitivity.}
As a complementary check, we additionally repeated the primary analysis under two alternative specifications of the priors on the $\Delta$ parameters: (b) substantially vaguer priors (standard deviations increased tenfold) and (c) mildly informative priors centred on small clinical expectations. The original priors are labeled (a). Table~\ref{tab:prior-sens} presents the posterior estimates for $\Delta E_0$ and $\Delta E_{\max}$ under all three specifications. The 95\% CrI for $\Delta E_0$ covers zero under each: $[-0.059, 0.013]$ under the original priors, $[-0.061, 0.005]$ under the vague (10$\times$ SD) priors, and $[-0.054, 0.002]$ under the informative priors. The posterior mean varies by at most 0.003 across specifications. For $\Delta E_{\max}$, the 95\% CrI covers zero under the original and vague priors but marginally excludes zero under the informative priors ($[0.010, 0.101]$), reflecting the pull of a prior centred on a small positive clinical expectation. The comparability conclusion for the immediate-response parameter $\Delta E_0$ is therefore insensitive both to tuning of the borrowing strength on the aggregate sources and to the tightness of the ethnic-comparability prior.

\begin{table}[ht!]
\centering
\caption{S5 --- Auxiliary ethnic-difference prior sensitivity: posterior means and 95\% CrIs for $\Delta E_0$ and $\Delta E_{\max}$ under three prior specifications for the $\Delta$ parameters (see Appendix~\ref{app:S5} for definitions of specifications (a)--(c)). All other model priors are unchanged.}
\label{tab:prior-sens}
\small
\begin{tabular}{lrlrl}
\toprule
Specification & $\Delta E_0$ & 95\% CrI & $\Delta E_{\max}$ & 95\% CrI \\
\midrule
(a) Original             & $-$0.0238 & [$-$0.0586,\ 0.0130] & 0.0458 & [$-$0.0223,\ 0.1124] \\
(b) Vague (10$\times$ SD) & $-$0.0266 & [$-$0.0610,\ 0.0050] & 0.0556 & [$-$0.0090,\ 0.1240] \\
(c) Informative          & $-$0.0270 & [$-$0.0540,\ 0.0020] & 0.0526 & [\ \ 0.0100,\ 0.1010] \\
\bottomrule
\end{tabular}
\end{table}

\subsubsection{S6 --- Tipping Point Analysis}

The leave-one-source-out analysis (S2) identified Publication~2 as the most influential single contributor to the location of $\Delta E_0$. To probe the robustness of the comparability conclusion to perturbations of that source, the clinical-score values from Publication~2 were systematically shifted by an additive constant $\delta$, and the model was refit for each shift. Because every ethnic-difference parameter has a 95\% credible interval that covers zero in the primary analysis, the relevant tipping point here is the smallest $|\delta|$ at which a perturbation causes any ethnic-difference parameter's 95\% credible interval to cross into excluding zero, thereby reversing the primary conclusion of ethnic comparability (see Appendix~\ref{app:S6} for the formal criterion).

Table~\ref{tab:tipping} presents results for shifts $\delta \in \{-0.20, -0.15, \ldots, +0.20\}$. At $\delta = 0$ every ethnic-difference parameter has a 95\% CrI covering zero, consistent with the primary comparability conclusion. The conclusion is fragile to downward perturbation of Publication~2 but robust to upward perturbation. A downward shift of only $\delta = -0.05$, the smallest step examined, already drives the 95\% CrI for $\Delta E_0$ below zero ($[-0.083, -0.002]$), reversing comparability in the immediate-response parameter, and $\Delta E_0$ remains below zero at every larger downward shift. At $\delta = -0.10$ the rebound parameter $\Delta E_{\max}$ also excludes zero ($[0.003, 0.208]$), so a moderate downward perturbation reverses comparability in both the immediate-response and the rebound-shape parameters. In the opposite direction no ethnic-difference parameter excludes zero at any upward shift up to $\delta = +0.20$. The drift parameter $\Delta b$ has a posterior mean on the order of $10^{-4}$ throughout and its 95\% CrI covers zero at every shift, so it never contributes to a tipping event. The tipping point for the comparability conclusion is therefore $\delta^\ast = -0.05$, and it is one-sided: the conclusion fails under a downward shift of as little as $0.05$ in the Publication~2 clinical scores but withstands upward shifts of up to $0.20$. The framework records this asymmetric fragility explicitly rather than reporting comparability as if it were unconditional.

\begin{table}[ht!]
\centering
\caption{S6 --- Tipping point analysis: posterior means and 95\% CrIs for the two primary ethnic-difference parameters ($\Delta E_0$ and $\Delta E_{\max}$) under additive shifts $\delta$ applied to all Publication~2 clinical-score values. The final column reports, per the formal tipping-point criterion (Appendix~\ref{app:S6}), all parameters in $\{\Delta E_0, \Delta E_{\max}, \Delta ED_{50}, \Delta r, \Delta b\}$ whose 95\% CrI excludes zero at each shift; numeric results for $\Delta ED_{50}$, $\Delta r$, and $\Delta b$ are omitted for compactness.}
\label{tab:tipping}
\small
\begin{tabular}{rrlrll}
\toprule
Shift $\delta$ & $\Delta E_0$ & 95\% CrI ($\Delta E_0$) & $\Delta E_{\max}$ & 95\% CrI ($\Delta E_{\max}$) & CrI excl.\ 0 \\
\midrule
$-$0.20 & $-$0.0962 & [$-$0.1558, $-$0.0388] & $\ \ $0.1063 & [$-$0.0034,\ 0.2256] & $\Delta E_0$ \\
$-$0.15 & $-$0.0755 & [$-$0.1368, $-$0.0047] & $\ \ $0.0850 & [$-$0.0457,\ 0.1952] & $\Delta E_0$ \\
$-$0.10 & $-$0.0717 & [$-$0.1287, $-$0.0209] & $\ \ $0.0947 & [\ \ 0.0027,\ 0.2079] & $\Delta E_0$, $\Delta E_{\max}$ \\
$-$0.05 & $-$0.0412 & [$-$0.0831, $-$0.0023] & $\ \ $0.0578 & [$-$0.0102,\ 0.1338] & $\Delta E_0$ \\
$\ $0.00 (primary) & $-$0.0238 & [$-$0.0586,\ \ 0.0130] & $\ \ $0.0458 & [$-$0.0223,\ 0.1124] & none \\
$+$0.05 & $-$0.0068 & [$-$0.0450,\ \ 0.0227] & $\ \ $0.0339 & [$-$0.0153,\ 0.1113] & none \\
$+$0.10 & $\ \ $0.0066 & [$-$0.0365,\ \ 0.0624] & $\ \ $0.0293 & [$-$0.0726,\ 0.1169] & none \\
$+$0.15 & $-$0.0024 & [$-$0.0589,\ \ 0.0334] & $\ \ $0.0688 & [$-$0.0030,\ 0.1706] & none \\
$+$0.20 & $\ \ $0.0421 & [$-$0.0257,\ \ 0.1236] & $\ \ $0.0009 & [$-$0.1612,\ 0.1126] & none \\
\bottomrule
\end{tabular}
\end{table}

\subsubsection{S7 --- Methodological Sensitivity}
\label{sec:s7-methods}

S7 asks whether replacing the primary borrowing mechanism with an alternative borrowing framework changes the conclusion. Whereas S5 perturbs the analyst-set controls within the primary framework (the power-prior weight $a_0$ and the priors on the $\Delta$ parameters), S7 substitutes a different mechanism for combining the external sources with the current-study data. A wide range of methods exists for this purpose, spanning Bayesian dynamic borrowing priors, frequentist test-then-pool procedures, propensity-score integration, doubly robust estimators, and recent selective borrowing approaches. Any of these could in principle serve as the alternative against which the primary analysis is compared. For the worked example we compare the primary analysis against two Bayesian frameworks that explicitly model between-source heterogeneity through a precision parameter: the commensurate prior and the robust meta-analytic-predictive (MAP) prior. Both share the primary model's likelihood and outcome structure, isolating the effect of the borrowing mechanism. Methods that fall outside the Bayesian family (e.g., test-then-pool, propensity-score integration) would require a different inferential framework, which would confound the borrowing-mechanism comparison with a change in uncertainty quantification. Therefore, they are not utilized here.

The two alternative frameworks are described below; complete mathematical specifications are provided in Appendix~\ref{app:borrowing-methods}.

\paragraph*{Commensurate prior.}
The commensurate prior \citep{Hobbs2011} introduces source-specific $E_0$ and $E_{\max}$ for the two aggregate sources with prior means centered at the Asian population parameters via a data-estimated commensurability precision $\tau_c$. Only Publication~1 and Publication~2 are subject to commensurate discounting.

\paragraph*{Robust MAP prior.}
The robust MAP prior \citep{Schmid2014} forms a predictive distribution for the current-study ethnic offsets from a hierarchy over the two aggregate sources and protects against prior-data conflict by incorporating a vague mixture component with weight $w_R=0.2$.

\medskip

Table~\ref{tab:borrowing-comparison} summarizes the key
trade-offs across the two alternative frameworks relative to the primary. Their posterior estimates of the ethnic-difference parameters in the worked example are compared in Table~\ref{tab:methods-compare}.

\begin{table}[ht!]
\centering
\caption{Trade-offs among the two alternative Bayesian borrowing frameworks used in the S7 analysis, with the primary (variance-scaled) analysis as reference.}
\label{tab:borrowing-comparison}
\small
\begin{tabular}{p{2.8cm} p{3.5cm} p{3.5cm} p{3.5cm}}
\toprule
 & Primary (variance-scaled) & Commensurate prior & Robust MAP prior \\
\midrule
Control parameter & Implicit $a_0 = 1$ (full incorporation) &
$\tau_c$, estimated & $\tau_{\mathrm{MAP}}$ estimated;
$w_R$ fixed \\
\addlinespace[3pt]
Adaptive to conflict? & No & Yes & Yes (via mixture) \\
\addlinespace[3pt]
Complexity & Low & Moderate & High \\
\addlinespace[3pt]
Best for & Simple full borrowing &
Single or few external sources with comparable structure &
Multiple exchangeable historical sources \\
\bottomrule
\end{tabular}
\end{table}

\noindent
The posterior means for $\Delta E_0$ are small and negative under all four configurations, ranging from $-0.024$ (primary) to $-0.032$ (robust MAP, current-study estimate). The 95\% CrI for $\Delta E_0$ covers zero under the primary model and continues to cover zero under the commensurate prior ($[-0.056, 0.0001]$, with the upper limit only marginally above zero) and both robust MAP summaries ($[-0.058, 0.011]$ for the population mean and $[-0.060, 0.009]$ for the current-study source-specific estimate). The 95\% CrI for $\Delta E_{\max}$ likewise covers zero in every framework. The comparability conclusion is therefore stable across the borrowing mechanism: frameworks that explicitly model between-source variability reach the same conclusion as the primary variance-scaled analysis. The commensurate prior estimates a between-source commensurability precision of $\hat{\tau}_c = 2.97$ (95\% CrI: $[0.63, 6.96]$), and the robust MAP prior estimates a between-source standard deviation of $\hat{\tau}_{\mathrm{MAP}} = 0.005$ (95\% CrI: $[0.0004, 0.017]$). Both point to limited residual heterogeneity in the ethnic-offset parameters once the model structure is accounted for, and the comparability conclusion holds whether or not that heterogeneity is modeled explicitly.

\begin{table}[ht!]
\centering
\caption{S7 --- Methodological sensitivity: posterior means and 95\% CrIs for $\Delta E_0$ and $\Delta E_{\max}$ under the two alternative Bayesian borrowing frameworks, with the primary analysis as reference. Robust MAP results are shown for the population mean (pop) and for the current-study source-specific estimate (CS).}
\label{tab:methods-compare}
\resizebox{\textwidth}{!}{
\begin{tabular}{llccccc}
\toprule
Framework & Configuration & $\Delta E_0$ & 95\% CrI & $\Delta E_{\max}$ & 95\% CrI \\
\midrule
Primary (variance-scaled) & $a_0 = 1.00$ & $-$0.0238 & [$-$0.0586,\ \ 0.0130] & 0.0458 & [$-$0.0223,\ 0.1124] \\
\addlinespace[3pt]
Commensurate prior & $\hat{\tau}_c = 2.97$ & $-$0.0272 & [$-$0.0559,\ \ 0.0001] & 0.0439 & [$-$0.0051,\ 0.0961] \\
\addlinespace[3pt]
Robust MAP prior & Pop.\ mean ($\Delta E_0^{\mathrm{pop}}$) & $-$0.0301 & [$-$0.0584,\ \ 0.0107] & 0.0566 & [$-$0.0191,\ 0.1097] \\
 & CS ($\Delta E_{0,1}$) & $-$0.0319 & [$-$0.0601,\ \ 0.0090] & 0.0557 & [$-$0.0202,\ 0.1090] \\
\bottomrule
\end{tabular}
}
\end{table}

\subsubsection{S8 --- Structural Model Sensitivity}

The primary analysis uses a two-component model (Emax plus linear trend). To test whether the conclusion depends on this functional form, two alternative models were fitted: (a)~an Emax-only model (dropping the linear trend), and (b)~a piecewise linear model with knots at Days~1, 28, and 84. Model comparison was performed using the Watanabe--Akaike Information Criterion (WAIC); see Appendix~\ref{app:S8} for model specifications and WAIC definitions.

Table~\ref{tab:waic} presents the WAIC comparison; Tables~\ref{tab:emax-only} and~\ref{tab:piecewise} report the posterior summaries for each alternative model. The primary Emax-plus-linear model achieves the lowest WAIC ($-3901.22$), ahead of the Emax-only model ($\Delta$WAIC $= 160.7$) and the piecewise linear model with three knots ($\Delta$WAIC $= 623.1$). The primary model is therefore preferred on predictive grounds by a wide margin, and it additionally carries direct mechanistic interpretation: an initial Emax response, a partial rebound captured by $E_{\max}$, and a long-term linear drift, each with an explicit ethnic offset. The piecewise model uses purely empirical segment slopes whose ethnic offsets do not map to standard pharmacodynamic quantities, and it fits substantially worse on this example. What matters for the credibility of the primary conclusion is whether the qualitative ethnic-comparability findings hold across structural specifications. In the primary model the 95\% CrI for $\Delta E_0$ covers zero. The Emax-only model departs from this on both ethnic-difference parameters: the 95\% CrI for $\Delta E_0$ narrowly excludes zero ($[-0.032, -0.003]$, mean $-0.018$) and the 95\% CrI for $\Delta E_{\max}$ also excludes zero ($[0.007, 0.039]$, mean $0.023$). Both departures are absent in the better-fitting primary model and are attributable to removing the linear-trend component: with no linear term to absorb the long-term drift, that drift is forced into the baseline-offset and rebound parameters, manufacturing apparent ethnic differences in $\Delta E_0$ and $\Delta E_{\max}$ rather than reflecting genuine ones. The much larger WAIC of the Emax-only model ($\Delta$WAIC $= 160.7$) confirms that these departures reflect model misspecification rather than a feature of the data. The piecewise model parameterises the early-time ethnic difference jointly through $\Delta E_0$ and the first-segment slope offset $\Delta\beta_1$, and the two are not separately identifiable from the single Day-1 visit covered by that segment (both posterior CrIs cover zero in Table~\ref{tab:piecewise}). What is identified is the combined Day-1 ethnic difference $\Delta E_0 + \Delta\beta_1 \cdot 1$, whose posterior mean is approximately $-0.03$, matching the primary $\Delta E_0$ estimate in both sign and magnitude. Among the later-time slope offsets, $\Delta\beta_2$ and $\Delta\beta_4$ have CrIs covering zero, while $\Delta\beta_3$ narrowly excludes zero ($[1\times10^{-4}, 8\times10^{-4}]$); the latter is on the order of $10^{-4}$ on the biomarker scale and carries no clinical interpretation. The comparability conclusion is therefore robust to structural choices in the mean function, with the only departures being marginal effects confined to the worse-fitting alternative models.

\begin{table}[ht!]
\centering
\caption{S8 --- WAIC model comparison across three structural formulations. $p_{\mathrm{WAIC}}$ is the effective number of parameters. $\Delta$WAIC is the difference from the best-fitting model.}
\label{tab:waic}
\small
\begin{tabular}{lrrr}
\toprule
Model & WAIC & $p_{\mathrm{WAIC}}$ & $\Delta$WAIC \\
\midrule
Primary (Emax + linear)      & $-$3901.22 & 11.01 &   0.0 \\
Emax-only                    & $-$3740.48 & 12.05 & 160.7 \\
Piecewise linear (3 knots)   & $-$3278.12 & 32.47 & 623.1 \\
\bottomrule
\end{tabular}
\end{table}

\begin{table}[ht!]
\centering
\caption{S8 --- Posterior summaries from the Emax-only model (no linear trend).}
\label{tab:emax-only}
\small
\begin{tabular}{lrrrrr}
\toprule
Parameter & Mean & Median & SD & 2.5\% & 97.5\% \\
\midrule
$E_0$            & $-$0.6421 & $-$0.6422 & 0.0064 & $-$0.6543 & $-$0.6294 \\
$\Delta E_0$     & $-$0.0175 & $-$0.0173 & 0.0075 & $-$0.0321 & $-$0.0034 \\
$E_{\max}$       & $\ \ $0.3336 & $\ \ $0.3338 & 0.0070 & $\ \ $0.3199 & $\ \ $0.3473 \\
$\Delta E_{\max}$& $\ \ $0.0229 & $\ \ $0.0228 & 0.0085 & $\ \ $0.0065 & $\ \ $0.0392 \\
$ED_{50}$        & $\ \ $6.860 & $\ \ $6.865 & 0.2698 & $\ \ $6.307 & $\ \ $7.384 \\
$\Delta ED_{50}$ & $\ \ $0.1445 & $\ \ $0.1506 & 0.3253 & $-$0.4919 & $\ \ $0.7667 \\
$r$              & $\ \ $2.128 & $\ \ $2.121 & 0.1671 & $\ \ $1.823 & $\ \ $2.477 \\
$\Delta r$       & $\ \ $0.0482 & $\ \ $0.0439 & 0.2467 & $-$0.4306 & $\ \ $0.5579 \\
$a$              & $-$0.0101 & $-$0.0101 & 2$\times 10^{-4}$ & $-$0.0106 & $-$0.0096 \\
$\sigma^2$       & $\ \ $0.0029 & $\ \ $0.0029 & 1$\times 10^{-4}$ & $\ \ $0.0027 & $\ \ $0.0031 \\
\bottomrule
\end{tabular}
\end{table}

\begin{table}[ht!]
\centering
\caption{S8 --- Posterior summaries from the piecewise linear model (knots at Days 1, 28, and 84). $\beta_1$--$\beta_4$ are the slope parameters for each time segment and $\Delta\beta_1$--$\Delta\beta_4$ are the corresponding ethnic offsets; see Appendix~\ref{app:S8} for definitions.}
\label{tab:piecewise}
\small
\begin{tabular}{lrrrrr}
\toprule
Parameter & Mean & Median & SD & 2.5\% & 97.5\% \\
\midrule
$E_0$           & $-$0.7726 & $-$0.7686 & 0.0933 & $-$0.9764 & $-$0.6244 \\
$\Delta E_0$    & $\ \ $0.0079 & $\ \ $0.0247 & 0.1583 & $-$0.3080 & $\ \ $0.2549 \\
$\beta_1$       & $\ \ $0.1988 & $\ \ $0.1945 & 0.0933 & $\ \ $0.0492 & $\ \ $0.4031 \\
$\Delta\beta_1$ & $-$0.0412 & $-$0.0590 & 0.1581 & $-$0.2893 & $\ \ $0.2747 \\
$\beta_2$       & $\ \ $0.0107 & $\ \ $0.0107 & 3$\times 10^{-4}$ & $\ \ $0.0100 & $\ \ $0.0113 \\
$\Delta\beta_2$ & $\ \ $5$\times 10^{-4}$ & $\ \ $5$\times 10^{-4}$ & 4$\times 10^{-4}$ & $-$4$\times 10^{-4}$ & $\ \ $0.0013 \\
$\beta_3$       & $\ \ $1$\times 10^{-4}$ & $\ \ $1$\times 10^{-4}$ & 2$\times 10^{-4}$ & $-$2$\times 10^{-4}$ & $\ \ $4$\times 10^{-4}$ \\
$\Delta\beta_3$ & $\ \ $4$\times 10^{-4}$ & $\ \ $4$\times 10^{-4}$ & 2$\times 10^{-4}$ & $\ \ $1$\times 10^{-4}$ & $\ \ $8$\times 10^{-4}$ \\
$\beta_4$       & $-$2$\times 10^{-4}$ & $-$2$\times 10^{-4}$ & $<$1$\times 10^{-4}$ & $-$3$\times 10^{-4}$ & $-$1$\times 10^{-4}$ \\
$\Delta\beta_4$ & $\ \ $0 & $\ \ $0 & $<$1$\times 10^{-4}$ & $-$1$\times 10^{-4}$ & $\ \ $1$\times 10^{-4}$ \\
$a$             & $-$0.0101 & $-$0.0101 & 3$\times 10^{-4}$ & $-$0.0107 & $-$0.0096 \\
$\sigma^2$      & $\ \ $0.0041 & $\ \ $0.0041 & 2$\times 10^{-4}$ & $\ \ $0.0038 & $\ \ $0.0044 \\
\bottomrule
\end{tabular}
\end{table}

\subsection{Summary of Sensitivity Findings}

Table~\ref{tab:sens-summary} provides a concise overview of all eight sensitivity analyses and their implications for the primary conclusion.
\begin{table}[htbp]
\centering
\footnotesize
\setlength{\tabcolsep}{3pt}
\caption{Summary of three-pillar sensitivity analysis results.}
\label{tab:sens-summary}
\begin{tabular}{@{}>{\raggedright\arraybackslash}p{0.6cm}
                >{\raggedright\arraybackslash}p{1.7cm}
                >{\raggedright\arraybackslash}p{7.2cm}
                >{\raggedright\arraybackslash}p{7.2cm}@{}}
\toprule
 & Analysis & Key Finding & Impact on Primary Conclusion \\
\midrule
\multicolumn{4}{l}{\textbf{Pillar 1: Appropriateness}} \\
\addlinespace[2pt]
& S1 (PPCs) & Close fit to dominant non-Asian source ($|z|<1.2$); the two aggregate publications show systematic biases in opposite directions (Publication~1 shallower, Publication~2 deeper), partially canceling in the joint posterior; small-sample Global Asian residuals are uniformly positive; RWE shows uniformly negative residuals with extreme $p$-values at several early-to-mid visits & Joint modeling is defensible; between-source heterogeneity in the aggregate sources motivates S2, S6, and S7; magnitude of aggregate-data residuals motivates S4 \\
\addlinespace[2pt]
& S2 (LOSO) & Publication~2 is the most influential single contributor: dropping it moves both $\Delta E_0$ ($[-0.081, -0.008]$) and $\Delta E_{\max}$ ($[0.017, 0.168]$) to exclude zero, reversing comparability; dropping any other single source leaves all CrIs covering zero & The comparability conclusion is contingent on Publication~2 rather than robust to dropping every source; Publication~2 is the appropriate target for the tipping-point probe in S6 \\
\midrule
\multicolumn{4}{l}{\textbf{Pillar 2: Value}} \\
\addlinespace[2pt]
& S3 (No-borrow) & Without aggregate data, the $\Delta E_0$ CrI excludes zero ($[-0.173, -0.001]$); borrowing pulls it back across zero into comparability; CrI for the Asian curve is $1.61\times$ wider at M6 and $1.30\times$ wider at M12 & Borrowing changes the inferential outcome (toward comparability), not just its precision \\
\addlinespace[2pt]
& S4 (ESS) & Publication~1: 45.6 patient-eq.\ at M6, 1.1 at M12 (collapses by M12); Publication~2: 25.7 (M6), 27.5 (M12); RWE: 44.9 (M6), 51.2 (M12) & RWE and Publication~2 sustain precision through M12; Publication~1's contribution is concentrated at M6 \\
\midrule
\multicolumn{4}{l}{\textbf{Pillar 3: Robustness}} \\
\addlinespace[2pt]
& S5 (Borrowing controls) & $\Delta E_0$ and $\Delta E_{\max}$ CrIs cover zero across power-prior weights $a_0 \in \{0.25, 0.50, 0.75, 1.00\}$ (CrI at $a_0 = 0.25$: $[-0.085, 0.007]$); under the three $\Delta$-prior specifications $\Delta E_0$ covers zero throughout (mean varies by $\leq 0.003$), while $\Delta E_{\max}$ covers zero except marginally under the informative prior & Comparability is robust to analyst tuning of borrowing strength and ethnic-comparability prior \\
\addlinespace[2pt]
& S6 (Tipping) & Comparability is fragile to downward perturbation of Publication~2 but robust to upward perturbation: a downward shift of as little as $\delta = -0.05$ makes $\Delta E_0$ exclude zero (and $\delta = -0.10$ also makes $\Delta E_{\max}$ exclude zero), whereas no ethnic-difference parameter excludes zero for upward shifts up to $\delta = +0.20$ & Comparability is not robust to downward perturbations of the most influential source; the framework records this conditionality explicitly \\
\addlinespace[2pt]
& S7 (Frameworks) & $\Delta E_0$ and $\Delta E_{\max}$ CrIs cover zero across all four configurations (Primary, Commensurate, MAP pop, MAP CS) & Comparability is not an artifact of the borrowing framework; explicit between-source heterogeneity modelling reaches the same conclusion \\
\addlinespace[2pt]
& S8 (Model) & Primary Emax+linear achieves best WAIC by a wide margin ($\Delta$WAIC $=160.7$ for Emax-only, $623.1$ for piecewise); $\Delta E_0$ CrI covers zero in the primary model and (jointly) in the piecewise model; the Emax-only model surfaces marginal $\Delta E_0$ and $\Delta E_{\max}$ differences attributable to dropping the linear trend & Comparability is robust to functional form; the only departures are marginal effects in the worse-fitting alternatives \\
\bottomrule
\end{tabular}
\end{table}

Applying the three-part conclusion template from Section~\ref{sec:framework}, an overall summary of the finding in Table~\ref{tab:sens-summary} is given below:
\subsubsection{Appropriateness (Pillar 1).}
The four data sources are sufficiently compatible to be jointly modeled within the hybrid framework. Posterior predictive checks (S1) show a close fit to the dominant non-Asian Global study and no structural misfit for the Asian IPD source, where the apparent residuals are an artefact of small per-visit samples. They do, however, identify meaningful forms of between-source heterogeneity that the framework is intended to detect and address: opposing systematic biases between the two aggregate sources (Publication~1 consistently shallower than predicted, residuals 0.016--0.023; Publication~2 consistently deeper, residuals $-$0.010 to $-$0.031) and uniformly negative residuals in the RWE source with extreme posterior-predictive $p$-values at several early-to-mid visits. Neither form disqualifies the joint fit. The joint model reconciles the opposing biases in the aggregate sources by averaging them, while the RWE deviation remains modest in magnitude. The leave-one-source-out analysis (S2) further confirms that Publication~2 exerts the strongest individual influence on the $\Delta E_0$ posterior, consistent with its long follow-up and uniformly negative residuals. This makes Publication~2 the natural focus for the tipping-point sensitivity analysis in S6 (Pillar 3). Overall, the observed heterogeneity is accommodated by the model, supporting the appropriateness of joint modeling.

\subsubsection{Value (Pillar 2).}
The aggregate external data materially change the inferential outcome rather than merely sharpening the estimates. Without Publication~1 and Publication~2, the 95\% CrI for $\Delta E_0$ excludes zero ($[-0.173, -0.001]$), pointing to a larger immediate response in Asian patients; adding the aggregate sources pulls the posterior back across zero, so the comparability conclusion depends on borrowing. The credible interval for the fitted Asian curve also becomes roughly $1.61\times$ wider at Month~6 and $1.30\times$ wider at Month~12 without borrowing. The effective sample size (ESS) decomposition in S4 shows that this added precision arises from complementary contributions: the RWE source anchors the late-time trajectory (ESS = 51.2 at Month~12), Publication~2 provides sustained precision at later visits (ESS = 27.5 at Month~12), and Publication~1 contributes strongly at Month~6 (ESS = 45.6) before its influence drops sharply once its data end. Hybrid evidence synthesis was therefore consequential in this setting: the regional IPD anchors the late-time trajectory, and the aggregate publications both shift the point estimate and supply the cross-sectional precision that resolves the ethnic-comparability question.

\subsubsection{Robustness (Pillar 3).}
The ethnic-comparability conclusion is robust along most dimensions but conditional on Publication~2. It is robust to borrowing strength within the primary framework (S5): the 95\% CrI for $\Delta E_0$ covers zero across the power-prior grid $a_0 \in \{0.25, 0.50, 0.75, 1.00\}$, including at the heaviest discount $a_0 = 0.25$ where the CrI is $[-0.085, 0.007]$, and under all three $\Delta$-prior specifications, with the posterior mean shifting by at most 0.003. It is robust to the choice of borrowing framework (S7): the commensurate prior and both robust MAP summaries leave every $\Delta$ CrI covering zero ($\hat{\tau}_c = 2.97$ for the commensurate prior; $\hat{\tau}_{\mathrm{MAP}} = 0.005$ for robust MAP), so explicitly modeling between-source heterogeneity reaches the same conclusion. It is largely robust to structural choices (S8): the primary Emax-plus-linear model attains the best WAIC by a wide margin, and the piecewise model's implied Day-1 ethnic difference $\Delta E_0 + \Delta\beta_1 \approx -0.03$ matches the primary estimate. The only departure is in the worse-fitting Emax-only model, where dropping the linear trend forces long-term drift into the baseline-offset and rebound terms, producing marginal $\Delta E_0$ and $\Delta E_{\max}$ differences absent in the better-fitting models. The conclusion is, however, conditional on Publication~2: dropping it (S2) moves both $\Delta E_0$ ($[-0.081, -0.008]$) and $\Delta E_{\max}$ ($[0.017, 0.168]$) to exclude zero, and the tipping-point analysis (S6) shows that comparability reverses under modest downward perturbations of that source. A shift of $\delta = -0.05$ in the Publication~2 clinical scores already makes $\Delta E_0$ exclude zero, whereas upward shifts up to $\delta = +0.20$ leave every $\Delta$ parameter covering zero. Ethnic comparability is therefore well supported by the assembled evidence and stable to how that evidence is weighted and modeled, but it rests materially on the longest-follow-up aggregate source. The analyses of the framework have made this dependence explicit.

\section{Discussion}
\label{sec:discussion}

\subsection{Lessons for Practitioners}

Application of the three-pillar framework to the worked example highlighted several practical lessons. Borrowing was consequential in this case: with only the IPD, the 95\% credible interval for $\Delta E_0$ excluded zero ($[-0.173, -0.001]$) and pointed to a larger immediate response in Asian patients, whereas adding the aggregate sources pulled the interval back across zero and supported ethnic comparability. The comparability conclusion therefore could not have been reached from the IPD alone. Teams facing sparse target-population data should include the no-borrowing reference (S3) to demonstrate how external information changes the conclusion.

The leave-one-source-out analysis (S2) identified Publication~2 as the most influential single contributor: dropping it moved both $\Delta E_0$ and $\Delta E_{\max}$ to exclude zero, reversing the comparability conclusion. Influential sources of this kind are the natural targets for tipping-point analysis (S6). For the worked example, comparability proved fragile to downward perturbation of Publication~2 but robust to upward perturbation: a downward shift of as little as 5 percentage points in its clinical-score values made $\Delta E_0$ exclude zero, whereas upward shifts of up to 20 percentage points left every ethnic-difference parameter covering zero. Teams should routinely pair S2 with S6 so that the dependence of a conclusion on dominant sources can be quantified.

Finally, the comparability conclusion was concordant across the four power-prior weights in S5 and the two alternative borrowing frameworks in S7: every $\Delta$ credible interval covered zero. This agreement is reassuring but does not mean unconditional robustness. The source-influence (S2) and tipping-point (S6) analyses show that the same conclusion depends on Publication~2 being included and unperturbed, where removing it or shifting it modestly reverses comparability. The informative contrast is therefore between two kinds of robustness: the conclusion is robust to how the evidence is weighted and modeled (S5, S7, S8) but not to the data sources themselves (S2, S6). Presenting all of these sensitivities alongside the primary analysis exposes this conditionality and allows reviewers to distinguish a conclusion supported by the full evidence base from the one that depends on a single influential source. We recommend that ECT submissions routinely pair borrowing-strength (S5) and borrowing-framework (S7) sensitivities with source-influence (S2) and tipping-point (S6) analyses, and disclose any source dependence explicitly.

\subsection{Complementarity with Methods Offering Internal-Validity Guarantees}

Modern borrowing methods increasingly carry strong internal guarantees. Adaptive Lasso Selective Borrowing \citep{Gao2025} discards external controls whose contribution to the loss function exceeds a data-driven threshold and provides asymptotic Type~I error control under adaptive-lasso oracle properties. Conformal Selective Borrowing \citep{Zhu2025} provides finite-sample exact Type~I error control through a randomization-inference framework and uses conformal $p$-values to drop external controls that fail a compatibility test. Doubly robust estimators \citep{Valancius2024} remain consistent if either the outcome model or the propensity model is correctly specified. Each method carries a guarantee that, conditional on its own modeling assumptions, the resulting estimate has the statistical properties advertised.

The three-pillar framework complements rather than replaces these borrowing methods. The two address different questions. Methods with internal-validity address one question: assuming the method’s assumptions hold, does the analysis perform as expected? The framework addresses a different question: regardless of the borrowing method used, are the decisions to borrow data appropriate, valuable, and robust to reasonable alternatives? These perspectives are complementary, and a defensible regulatory submission needs both. We recommend pairing any primary borrowing method—chosen for its statistical properties and suitability to the study—with the structured evaluation provided by the three pillars. The primary method supplies the core inference; the framework provides the credibility assessment.

\subsection{Limitations}
The framework is descriptive rather than decision-theoretic. It organizes the questions a sensitivity analysis should answer but does not specify quantitative thresholds (for example, the magnitude of tipping point that qualifies as robust) nor a formal rule for combining evidence when individual analyses show discordant signals. Whether borrowing-control sensitivity (S5) should weaken confidence in an otherwise concordant pattern across alternative borrowing methods (S7) and structural models (S8), or whether a small tipping point (S6) overrides agreement on the no-borrowing reference (S3), remains a matter of contextual judgment by the study team and reviewers.

The assessment of borrowing in the framework operates at the level of statistical compatibility and cannot diagnose causal exchangeability of data sources. Per-source predictive checks (S1) identify when external sources disagree distributionally with the current study or with one another, but they cannot determine whether external real-world cohorts share the same treatment effect mechanism, namely whether the same confounders are balanced and the same selection processes apply. A pass on heterogeneity diagnostics is therefore evidence of statistical compatibility, not of causal exchangeability.

The worked example has features that simplified application of the framework: a continuous outcome suitable for parametric modeling, multiple external sources permitting per-source diagnostics and leave-one-out analyses, and ethnic-difference parameters that decompose cleanly within an Emax model. Settings with binary or time-to-event endpoints, a single external source, or non-parametric primary analyses may require additional adaptation, especially of the multi-source and parametric-likelihood analyses (S1, S2, S4). 

\subsection{Conclusion}

This paper contributes a practical three-pillar sensitivity analysis framework for evaluating information borrowing in ECTs and hybrid evidence synthesis. The framework is designed to be method-agnostic and is illustrated in detail through a simulated worked example that reflects the hybrid evidence packages now appearing under regulatory real-world-evidence pathways. It operationalizes currently implicit regulatory expectations for sensitivity analysis and provides a structured, reproducible approach that can be applied to Bayesian dynamic borrowing as well as to frequentist and causal-inference methods such as propensity-score augmentation and target-trial emulation. As regulatory acceptance of external evidence expands, structured sensitivity analysis will become increasingly important in submissions and will serve as a natural complement to the internal-validity properties of any specific borrowing method.

\section{Software and Reproducibility}

All analyses were conducted using \texttt{nimble} in \texttt{R} \citep{deValpine2017}. The data used in the worked example are simulated. All models were fitted with two MCMC chains; each chain discarded 10{,}000 run-in iterations and then retained 100{,}000 post-run-in draws after thinning by a factor of five, giving 200{,}000 posterior draws in total. MCMC convergence was assessed using trace plots, the Gelman--Rubin $\hat{R}$ statistic, and effective sampling draw calculations; full per-parameter diagnostics for the primary analysis are reported in Table~\ref{tab:convergence}. All $\hat{R}$ point estimates are at or below 1.002 with upper confidence bounds at or below 1.010, and every monitored parameter has an effective sampling draw above 1{,}500. All R code and simulated datasets are publicly available at 
https://github.com/gux9/ect-sensitivity-framework.

\section{Disclosure Statement}\label{disclosure-statement}
This research received no funding from any public, commercial, or not-for-profit agency. The authors have no conflicts of interest to declare. 

To assist with language editing and coding tasks, the authors used Opus 4.7 from Anthropic during manuscript preparation. The tool was not used to generate research data, statistical results, or figures. All AI-assisted output was reviewed by the authors, who take full responsibility for the content of the manuscript.

\bibliography{Three_pillar_bibliography.bib}

%
%

\newpage
\appendix
\setcounter{table}{0}
\renewcommand{\thetable}{A\arabic{table}}
\setcounter{figure}{0}
\renewcommand{\thefigure}{A\arabic{figure}}
\setcounter{equation}{0}
\renewcommand{\theequation}{A.\arabic{equation}}

\begin{center}
{\LARGE\bf Supplementary Appendix}
\end{center}
\bigskip

This appendix provides the complete mathematical specifications for all models and sensitivity analyses described in the main text, including notation conventions and derivations.

\section{Sensitivity Analysis Methods}
\label{app:sensitivity}

\subsection{Posterior Predictive Checks (S1)}
\label{app:S1}

For each data source $s$ and visit day $d$, the posterior predictive distribution of the mean response is $\hat{\mu}_{sd}^{(m)} = f(x_d, z_s, I_s; \btheta^{(m)})$ \citep{Gelman1996PPC}. The diagnostics are:

\begin{tabular}{lp{9.5cm}}
\toprule
Statistic & Formula \\
\midrule
Predicted mean & $\bar{\hat{\mu}}_{sd} = M^{-1}\sum_m \hat{\mu}_{sd}^{(m)}$ \\[3pt]
Residual & $r_{sd} = y_{sd}^{\mathrm{obs}} - \bar{\hat{\mu}}_{sd}$ \\[3pt]
PPC $p$-value & $p_{sd} = M^{-1}\sum_m \mathbf{1}(\hat{\mu}_{sd}^{(m)} \le y_{sd}^{\mathrm{obs}})$; values near 1 indicate the model under-predicts at that cell (observed exceeds most replicates), values near 0 indicate over-prediction (observed falls below most replicates), and values near 0.5 indicate adequate fit \\
\bottomrule
\end{tabular}

\subsection{Leave-One-Source-Out (S2)}
\label{app:S2}

For each Asian source $s$, refit the primary model on the reduced dataset $\mathcal{D} \setminus \mathcal{D}_s$. The baseline biomarker centering is recomputed per reduced dataset: $\bar{z}_w^{(-s)} = \sum_{i \notin \mathcal{D}_s} n_i z_i^{\mathrm{raw}} \big/ \sum_{i \notin \mathcal{D}_s} n_i$. The posterior means and 95\% CrIs of all $\Delta$ parameters are compared across the full and four leave-one-out fits to identify the most influential source.

\subsection{No-Borrowing Reference (S3)}
\label{app:S3}

Fit the primary model using only $\mathcal{D}_{\mathrm{IPD}} = \{i : n_i = 1\}$ (Global study + RWE). The 95\% CrI width for the Asian curve at reference time points (Day~168, Day~336) is compared:
\begin{equation}
 W_d = Q_{0.975}\!\left(\hat{\mu}_d^{\text{Asian}}\right) 
 - Q_{0.025}\!\left(\hat{\mu}_d^{\text{Asian}}\right), 
 \qquad
 \text{Width ratio} = \frac{W_d^{\text{no-borrow}}}{W_d^{\text{full}}}.
\end{equation}

\subsection{Effective Sample Size Decomposition (S4)}
\label{app:S4}

Following the variance-ratio approach of \citet{Morita2008ESS}, let $V_{\mathrm{full}}$ be the posterior variance of the fitted Asian curve at a reference time point from the full model, $V^{(-s)}$ from the LOSO analysis dropping source $s$, and $V_{\mathrm{IPD}}$ from the no-borrowing analysis. The ESS of source $s$ is:
\begin{equation}
 \mathrm{ESS}_s 
 = \frac{1/V_{\mathrm{full}} - 1/V^{(-s)}}{\frac{1}{V_{\mathrm{IPD}}} \big/ N_{\mathrm{Asian\;IPD}}}
 = \left(\frac{1}{V_{\mathrm{full}}} - \frac{1}{V^{(-s)}}\right)
 \cdot V_{\mathrm{IPD}} \cdot N_{\mathrm{Asian\;IPD}},
\end{equation}
where $N_{\mathrm{Asian\;IPD}} = 60$ (10 Global study + 50 RWE). The supplementary variance ratio $\mathrm{VarRatio}_s = V^{(-s)}/V_{\mathrm{full}}$ is also reported (values ${>}1$ confirm precision contribution).

\subsection{Borrowing-Control Sensitivity (S5)}
\label{app:S5}

S5 comprises two sub-analyses: (i) a power-prior weight grid that varies the borrowing strength applied to the two aggregate sources within the primary borrowing family, and (ii) a sensitivity check on the priors of the ethnic-difference parameters $\Delta$.

\subsubsection{Power-prior weight grid.} The aggregate-data likelihood for Publication~1 and Publication~2 is raised to a power $a_0 \in \{0.25, 0.50, 0.75, 1.00\}$, with the RWE study retaining full weight throughout. The configuration $a_0 = 1$ recovers the primary analysis. The formal specification is given in Appendix~\ref{app:power-prior}.

\subsubsection{Ethnic-difference prior specifications.} The primary model is refit under three prior specifications for the $\Delta$ parameters. Specification~(a) uses the original priors given in Eq.~\eqref{eq:prior}. Specifications~(b) and~(c) modify only the $\Delta$-parameter priors; all other priors remain unchanged.

\begin{center}
\small
\begin{tabular}{lp{9cm}}
\toprule
Specification & Modification to $\Delta$-parameter priors \\
\midrule
(a) Original & As specified in Eq.~\eqref{eq:prior}. \\[4pt]
(b) Vague & Prior variance of each $\Delta$ parameter multiplied by a factor of 100 (i.e., prior SD multiplied by 10). For example, $\Delta E_0 \sim \mathcal{N}(0, 100)$ instead of $\mathcal{N}(0, 1)$; $\Delta b \sim \mathcal{N}(0, 1000)$ instead of $\mathcal{N}(0, 10)$. \\[4pt]
(c) Informative & Prior SD of each $\Delta$ parameter halved relative to the original, with means shifted to small clinically plausible offsets. For example, $\Delta E_0 \sim \mathcal{N}(-0.05,\, 0.25)$ instead of $\mathcal{N}(0, 1)$; $\Delta E_{\max} \sim \mathcal{N}(0.05,\, 0.25)$ instead of $\mathcal{N}(0, 1)$. \\
\bottomrule
\end{tabular}
\end{center}

\subsection{Tipping Point Analysis (S6)}
\label{app:S6}

Tipping point analysis adapts an idea from sensitivity analysis for missing data \citep{Daniels2008Missing, Liublinska2014Tipping} to the external-borrowing setting. For a grid of shifts $\delta$, create $\tilde{y}_{i}^{\mathrm{P2}} = y_{i}^{\mathrm{P2}} + \delta$ for all Publication~2 data points, refit the primary model, and compute posterior CrIs for all $\Delta$ parameters. For each parameter $\theta$ and shift $\delta$, define the CrI-exclusion indicator
\begin{equation}
 E(\theta,\delta) = \mathbbm{1}\!\left\{Q_{0.025}(\theta|\delta) > 0 \text{ or } Q_{0.975}(\theta|\delta) < 0\right\},
\end{equation}
which equals $1$ when the 95\% CrI excludes zero and $0$ otherwise. The tipping point is the smallest absolute shift at which the exclusion state of at least one $\Delta$ parameter changes relative to the primary analysis $(\delta = 0)$:
\begin{equation}
 \delta^* = \min\!\left\{|\delta| : \exists\;\theta \in 
 \{\Delta E_0, \Delta E_{\max}, \Delta ED_{50}, \Delta r, \Delta b\}
 \text{ s.t.\ } E(\theta,\delta) \neq E(\theta,0)
 \right\}.
\end{equation}
This criterion captures a reversal in either direction: a parameter whose CrI excludes zero at $\delta = 0$ is tipped when its CrI first comes to cover zero, and a parameter whose CrI covers zero at $\delta = 0$ (such as $\Delta E_0$) is tipped when its CrI first comes to exclude zero.

\section{Alternative Bayesian Borrowing Methods}
\label{app:borrowing-methods}

Let $\btheta$ denote the shared model parameters, $D$ the current-study data, and $D_h$ the data from the $h$-th external source. This appendix documents three Bayesian borrowing methods used in the sensitivity analyses. The power prior (Section~\ref{app:power-prior}) is used in S5 to vary the borrowing strength within the primary borrowing family; the commensurate prior (Section~\ref{app:commensurate}) and the robust meta-analytic-predictive prior (Section~\ref{app:map}) are used in S7 as alternative borrowing frameworks.

\subsection{Power Prior}
\label{app:power-prior}

The power prior \citep{Ibrahim2015} discounts external data by raising the aggregate-data likelihood to a power $a_0 \in [0,1]$. In this implementation, the data are split into three blocks:

\subsubsection{Global study IPD ($i = 1,\ldots,N_{\mathrm{IPD}}$):}
$y_i \sim \mathcal{N}(\mu_i, \sigma^2)$, where $\mu_i$ is the primary mean function~\eqref{eq:XEN-mean}.

\subsubsection{RWE ($j = 1,\ldots,N_{\mathrm{RWE}}$), full weight:} 
$y_j^{\mathrm{RWE}} \sim \mathcal{N}(\mu_j^{\mathrm{RWE}},\; \sigma^2)$.

\subsubsection{Publication 1 / Publication 2 ($m = 1,\ldots,N_{\mathrm{disc}}$), discounted:}
\begin{equation}
 y_m^{\mathrm{disc}} \sim \mathcal{N}\!\left(\mu_m^{\mathrm{disc}},\;
 \frac{\sigma^2}{n_m \cdot a_0}\right).
 \label{eq:app-pp}
\end{equation}

\noindent
Equation~\eqref{eq:app-pp} is equivalent to raising the aggregate-data likelihood to the power $a_0$: $L(\btheta \mid \mathcal{D}_{\mathrm{disc}})^{a_0} \propto \exp\!\left(-\frac{a_0}{2\sigma^2}\sum_m n_m(y_m - \mu_m)^2\right)$.

\subsection{Commensurate Prior}
\label{app:commensurate}

The commensurate prior \citep{Hobbs2011} introduces source-specific parameters for the intercept and maximum rebound in each external Asian source, linked to the current-study Asian parameters via a commensurability precision.

\subsubsection{IPD block.} Same as the primary model.

\subsubsection{Aggregate block.} For observation $j$ from external source 
$s(j)$:
\begin{equation}
 \mu_j^{\mathrm{agg}} =
 E_{0,s(j)}^{\mathrm{agg}} + a\,z_j
 + (b + \Delta b)\,x_j
 + \frac{E_{\max,s(j)}^{\mathrm{agg}}\; x_j^{r+\Delta r}}
 {(ED_{50}+\Delta ED_{50})^{r+\Delta r} + x_j^{r+\Delta r}}.
\end{equation}

\subsubsection{Commensurate priors (centred on Asian parameters).}
\begin{equation}
 E_{0,s}^{\mathrm{agg}} \sim \mathcal{N}(E_0 + \Delta E_0,\; \tau_c^{-1}),
 \qquad
 E_{\max,s}^{\mathrm{agg}} \sim \mathcal{N}(E_{\max} + \Delta E_{\max},\; \tau_c^{-1}),
 \qquad
 \tau_c \sim \mathrm{Gamma}(1, 1).
 \label{eq:app-comm}
\end{equation}

\noindent
Note: Because $E_{0,s}^{\mathrm{agg}}$ is centred on the Asian intercept $E_0 + \Delta E_0$, the ethnic offset $\Delta E_0$ is not added separately in $\mu_j^{\mathrm{agg}}$; doing so would double-count it. The same applies to $E_{\max,s}^{\mathrm{agg}}$ and $\Delta E_{\max}$.

\subsection{Robust Meta-Analytic-Predictive (MAP) Prior}
\label{app:map}

The robust MAP prior \citep{Schmid2014} derives a predictive distribution for the current-study parameters from a hierarchical model across external sources. 

\subsubsection{Source indexing.} Let $s = 1, 2$ index the two aggregate Asian sources ($s = 1$: Publication~1; $s = 2$: Publication~2). The current-study ethnic offsets are denoted without a source subscript ($\Delta E_0$, $\Delta E_{\max}$).

\subsubsection{MAP hierarchy across the two aggregate sources:}
\begin{equation}
 \Delta E_{0,s} \sim \mathcal{N}(\Delta E_0^{\mathrm{pop}},\; \tau_{\mathrm{MAP}}^2),
 \qquad
 \Delta E_{\max,s} \sim \mathcal{N}(\Delta E_{\max}^{\mathrm{pop}},\; \tau_{\mathrm{MAP}}^2),
 \quad s = 1, 2.
\end{equation}

\subsubsection{Robust mixture prior for the current-study offsets:}
\begin{equation}
 \Delta E_0 \sim (1-w_R)\;\mathcal{N}(\Delta E_0^{\mathrm{pop}},\;\tau_{\mathrm{MAP}}^2)
 + w_R\;\mathcal{N}(0,\;\sigma_{\mathrm{vague}}^2),
\end{equation}
with $w_R = 0.2$ and $\sigma_{\mathrm{vague}} = 10$. An analogous robust mixture is used for $\Delta E_{\max}$. The Global Asian subgroup and the RWE study contribute through the primary likelihood and are not part of the MAP hierarchy.

\subsubsection{Hyperpriors:}
$\Delta E_0^{\mathrm{pop}} \sim \mathcal{N}(0, 1)$,\quad
$\Delta E_{\max}^{\mathrm{pop}} \sim \mathcal{N}(0, 1)$,\quad
$\tau_{\mathrm{MAP}} \sim \mathcal{N}^+(0, 0.25)$.

\subsection{Connections Across Frameworks}

Although the three borrowing methods differ in parameterisation, they share a common structure: each places the degree of borrowing on a continuum between full incorporation and complete discounting of historical data. The power prior controls borrowing through $a_0$; the commensurate prior through $\tau_c$; and the MAP prior through $\tau_{\mathrm{MAP}}$ and $w_R$. The primary analysis uses the variance scaling $\sigma^2/n_i$, which is equivalent to fixing $a_0 = 1$ for all sources in the power-prior family. The S5 borrowing-control sensitivity moves within that family by varying $a_0$, while the S7 methodological sensitivity moves across families by substituting the commensurate or robust MAP prior. Concordant conclusions across the within-family variation (S5) speak to the choice of borrowing strength; concordant conclusions across the across-family substitutions (S7) speak to the choice of borrowing framework.

\section{Structural Model Alternatives (S8)}
\label{app:S8}

\subsection{Emax-Only Model}

Drops the linear trend from the mean function:
\begin{equation}
 \mu_{i} = (E_0 + \Delta E_0 \cdot I_i) + a\,z_i
 + \frac{(E_{\max}+\Delta E_{\max}\cdot I_i)\;x_{d_i}^{\,r+\Delta r\cdot I_i}}
 {(ED_{50}+\Delta ED_{50}\cdot I_i)^{r+\Delta r\cdot I_i}
 + x_{d_i}^{\,r+\Delta r\cdot I_i}}.
\end{equation}

\subsection{Piecewise Linear Model}

Four segments with fixed knots at Days 1, 28, and 84:
\begin{equation}
 \mu_{i} = (E_0 + \Delta E_0 \cdot I_i) + a\,z_i + \mathrm{PW}(x_d, I_i),
\end{equation}
where
\begin{equation}
 \mathrm{PW}(x, I) = \begin{cases}
 (\beta_1 + \Delta\beta_1 I)\,x
 & 0 < x \le 1 \\[2pt]
 V_1 + (\beta_2 + \Delta\beta_2 I)(x - 1)
 & 1 < x \le 28 \\[2pt]
 V_2 + (\beta_3 + \Delta\beta_3 I)(x - 28)
 & 28 < x \le 84 \\[2pt]
 V_3 + (\beta_4 + \Delta\beta_4 I)(x - 84)
 & x > 84
 \end{cases}
\end{equation}
with continuity conditions: $V_1 = (\beta_1 + \Delta\beta_1 I) \cdot 1$,\; $V_2 = V_1 + (\beta_2 + \Delta\beta_2 I)(28 - 1)$,\; $V_3 = V_2 + (\beta_3 + \Delta\beta_3 I)(84 - 28)$.

\subsection{WAIC for Model Comparison}

Model comparison is performed using the Watanabe--Akaike Information Criterion \citep{Watanabe2010WAIC, Gelman2014WAIC}:
\begin{equation}
 \mathrm{WAIC} = -2\,\widehat{\mathrm{lppd}} + 2\,p_{\mathrm{WAIC}},
\end{equation}
where $\widehat{\mathrm{lppd}} = \sum_{i=1}^N \log\!\left(M^{-1}\sum_m p(y_i \mid \btheta^{(m)})\right)$ and $p_{\mathrm{WAIC}} = \sum_{i=1}^N \mathrm{Var}_m\!\left(\log p(y_i \mid \btheta^{(m)})\right)$. Lower WAIC is preferred. Models within approximately 2--4 units are considered practically equivalent.

\section{Posterior Summary Statistics}
\label{app:summaries}

All MCMC analyses produce posterior samples 
$\{\btheta^{(1)},\ldots,\btheta^{(M)}\}$ from which:
\begin{table}
\begin{center}
\begin{tabular}{ll}
\toprule
Statistic & Formula \\
\midrule
Posterior mean & $\bar{\theta} = M^{-1}\sum_m \theta^{(m)}$ \\[4pt]
Posterior median & $Q_{0.5}(\theta)$ \\[4pt]
Posterior SD & $\mathrm{SD} = \sqrt{(M{-}1)^{-1}\sum_m(\theta^{(m)}-\bar{\theta})^2}$ \\[4pt]
95\% CrI & $[Q_{0.025},\; Q_{0.975}]$ \\
\bottomrule
\end{tabular}
\end{center}
\end{table}

\noindent
\textbf{Interpretation for ethnic comparability:} For each $\Delta$ parameter, a posterior mean near zero with a 95\% CrI that covers zero indicates the data are consistent with no ethnic difference. A narrow CrI centered on zero provides strong evidence for comparability; a wide CrI covering zero indicates insufficient precision to distinguish the ethnic difference from zero.

\subsection{Convergence Diagnostics}

The Gelman--Rubin $\hat{R}$ statistic \citep{GelmanRubin1992} compares within-chain and between-chain variance ($\hat{R} < 1.05$ indicates convergence). The effective sampling draw (ESD) accounts for autocorrelation within chains; values above 100 per chain generally ensure reliable posterior summaries.

Table~\ref{tab:convergence} reports the Gelman--Rubin $\hat{R}$ statistics and MCMC effective sampling draws for the primary analysis ($\mathrm{InvGamma}(0.01, 0.01)$ prior). All $\hat{R}$ point estimates are at or below 1.002 and all upper confidence bounds are at or below 1.010, well within the $\hat{R} < 1.05$ convergence threshold. Effective sampling draws range from 1{,}589 (for $\Delta E_{\max}$) up to 201{,}186 (for the fast-mixing variance parameters $a$ and $\sigma^2$), and every monitored parameter exceeds the 200 sample reliability threshold for two chains. To attain these effective sampling draws for the more slowly mixing Emax-block and ethnic-difference parameters, the primary analysis employed a block sampler targeting the four correlated Emax parameters $(E_0, \Delta E_0, E_{\max}, \Delta E_{\max})$ together with slice samplers for the Hill-coefficient parameters $(r, \Delta r)$, in addition to standard random-walk updates for the remaining parameters. The credible intervals underlying the primary comparability conclusions are therefore estimated with adequate Monte Carlo precision.

\begin{table}[ht!]
\centering
\caption{Convergence diagnostics for the primary analysis MCMC ($\mathrm{InvGamma}(0.01, 0.01)$ prior on $\sigma^2$). $\hat{R}$: Gelman--Rubin statistic (point estimate and upper 97.5\% CI). ESD: effective sampling draw.}
\label{tab:convergence}
\small
\begin{tabular}{lrrr}
\toprule
Parameter & $\hat{R}$ (point est.) & $\hat{R}$ (upper CI) & ESD \\
\midrule
$a$ & 1.000 & 1.000 & 201\,186 \\
$\Delta E_0$ & 1.001 & 1.006 & $\ \ $1\,918 \\
$\Delta ED_{50}$ & 1.000 & 1.001 & $\ \ $4\,732 \\
$\Delta E_{\max}$& 1.002 & 1.010 & $\ \ $1\,589 \\
$\Delta b$ & 1.002 & 1.009 & $\ \ $2\,064 \\
$\Delta r$ & 1.001 & 1.007 & $\ \ $1\,898 \\
$E_0$ & 1.001 & 1.002 & $\ \ $2\,114 \\
$ED_{50}$ & 1.000 & 1.002 & $\ \ $4\,852 \\
$E_{\max}$ & 1.002 & 1.005 & $\ \ $1\,750 \\
$b$ & 1.002 & 1.007 & $\ \ $2\,205 \\
$r$ & 1.001 & 1.003 & $\ \ $2\,154 \\
$\sigma^2$ & 1.000 & 1.000 & 195\,137 \\
\bottomrule
\end{tabular}
\end{table}

\subsection{Computation of Posterior Predictive Curves}
\label{app:fitted-curves}

For each MCMC draw $\btheta^{(m)}$, $m=1,\dots,M$, the fitted curve is evaluated at the study-visit days $\{x_d\}_{d=1}^D$:
\begin{align}
 \hat{\mu}_d^{\text{Non-Asian},(m)} &= 
 f\!\left(x_d,\; \bar{z}^{\text{NA}},\; I=0;\; \btheta^{(m)}\right), \\
 \hat{\mu}_d^{\text{Asian},(m)} &= 
 f\!\left(x_d,\; \bar{z}^{\text{A}},\; I=1;\; \btheta^{(m)}\right),
\end{align}
where $f(\cdot)$ is the mean function~\eqref{eq:XEN-mean}, $\bar{z}^{\text{NA}}$ is the mean centered baseline biomarker for non-Asian patients, and $\bar{z}^{\text{A}}$ for Asian patients. The posterior mean curve and 95\% credible band at each day are
\begin{equation}
 \bar{\hat{\mu}}_d = \frac{1}{M}\sum_{m=1}^M \hat{\mu}_d^{(m)},
 \qquad
 \CrI_d = \left[Q_{0.025}\!\left(\{\hat{\mu}_d^{(m)}\}_{m=1}^M\right),\;
 Q_{0.975}\!\left(\{\hat{\mu}_d^{(m)}\}_{m=1}^M\right)\right].
\end{equation}
Overlapping Asian and non-Asian credible bands indicate ethnic comparability in the biomarker response over time.

\end{document}